\documentclass[aps,prl,reprint,groupedaddress]{revtex4-2}
\usepackage{dcolumn}
\usepackage{bm}
\usepackage[utf8]{inputenc}
\usepackage[T1]{fontenc}
\usepackage{booktabs, array, mathptmx, float, tabularx, booktabs, lipsum, amsmath,multirow,amssymb}
\usepackage{siunitx, xcolor}
\usepackage[version=4]{mhchem}
\usepackage{graphicx,amssymb}
\usepackage{subfigure}
\usepackage{indentfirst,latexsym,bm}
\usepackage{threeparttable}
\graphicspath{{figs/}{figsgaoerb/}}
\usepackage[colorlinks,linkcolor=blue,anchorcolor=blue,citecolor=blue]{hyperref}
\usepackage{hyperref}
\usepackage{amsfonts}
\usepackage{newtxtext,newtxmath}

\begin{document}

\title{From orthogonal link to phase vortex in generalized dynamical Hopf insulators}
\author{Yuxuan Ma$^{1,2}$}
\author{Xin Li$^{1}$}
\email{To whom correspondence should be addressed. Email: lixin$_$physics@126.com}
\author{Yu Wang$^{1}$}
\author{Shuncai Zhao$^{1}$}
\author{Guangqin Xiong$^{1}$}
\author{Tongxin Sun$^{1}$}
\affiliation{$^{1}$Data science research center, Kunming University of Science and Technology, Kunming, 650093, China}
\affiliation{$^{2}$Department of Physics, Shanghai Normal University, Shanghai 200234, China}

\begin{abstract}
	
	 In the creation
	of Hopf topological matters, the old paradigm is to conceive the Hopf invariant first, and then display its intuitive
	topology through links. Here we brush aside this effort and put forward a new recipe for unraveling the
	 quenched  two-dimensional ($2D$) two-band Chern insulators  under a parallel quench protocol, which implies that the quench quantities with different momentum $\bm{k}$ are parallel or antiparallel to each other. We find that whether the dynamical Hopf invariant exists or not,  the links in $(2+1)D$ space always keep their standard shape even for topological initial states, and trace out the trajectories of phase vortices. The linking number is exactly equal to the difference between pre- and post-quench Chern numbers regardless of the construction of homotopy groups. We employ two concrete examples to illustrate these results, highlighting the polarity reversal at fixed points.
\end{abstract}
\maketitle
\section{\uppercase\expandafter{\romannumeral1}.~~INTRODUCTION}
Recent years have seen a tremendous surge in research of topological quantum matters attributed to not only their own striking natures \cite{hasan2010colloquium, asboth2016short,luo2019advanced} but also the potential applications ranging from quantum computation to metrology and spintronics \cite{stanescu2016introduction,zeng2019quantum,he2019topological, yu2022experimental}. In theoretical terms, the tenfold periodic table has served as a cornerstone
for categorizing non-interacting fermionic topological insulators and superconductors into topological equivalence classes \cite{schnyder2008classification, kitaev2009periodic,ryu2010topological}. A peculiar exception beyond the standard classification paradigm is the Hopf insulator \cite{moore2008topological,deng2013hopf,schuster2019floquet,lapierre2021n}, first confirmed experimentally in a single nitrogen-vacancy center \cite{yuan2017observation}, then in circuit networks \cite{wang2023realization}.

 Further more, the exploration of topological properties far from equilibrium has also grown into a major endeavor due to the rapid advancements of novel manipulating techniques \cite{rudner2020band,eisert2015quantum, meier2016observation,reid2022dynamically, jotzu2014experimental, wu2016realization, cooper2019topological, eckardt2017colloquium, gross2017quantum, schafer2020tools}. Two typical strategies for pulling a system out of equilibrium are periodic driving and sudden quenching. Relatedly, the nonequilibrium topological classifications have taken effect \cite{zhang2018dynamical, mcginley2019classification}; The Hopf invariant has also extended to Non-Hermitian \cite{he2020non}, Euler class \cite{unal2020topological},  and floquet  systems \cite{unal2019hopf}, etc. In addition, a particular attention has been devoted to two-dimensional ($2D$) two-band Chern insulators after quench, i.e, so-called dynamical Hopf insulators, where a link in $(2+1)D$ momentum-time space provides direct access to the static band topology \cite{wang2017scheme, chen2020linking}.

The link has first been used to explain the origin of stable matter by Lord Kelvin  more than a century ago, then become an obsession, theoretically and experimentally; The vortex as $2D$ topological defect has attracted a great deal of interest since early times,  being central to superfluidity, superconductivity, wave-guides and  Bose–Einstein condensation in quantum physics. The synergy of the two different concepts deepened the depiction of the dynamic topology only limited in a special quenching pattern starting from  topologically trivial initial states \cite{tarnowski2019measuring,yu2017phase}, but several intrinsic features still remain shrouded in mystery, the hidden mechanisms in dynamical Hopf insulators expect to be grasped.

To address this issue, let us briefly recap the main thread before starting our work. Here we focus on the sudden quench of $2D$ two-band Chern insulators by preparating an initial ground state $|\psi(\bm{k},0)\rangle$ of $\mathcal{H}^{i}$ at $t=0$: $\mathcal{H}^{i}=\frac{1}{2}\bm{H}^{i}(\bm{k})\cdot\bm{\sigma}\rightarrow \mathcal{H}^{f}=\frac{1}{2}\bm{H}^{f}(\bm{k})\cdot\bm{\sigma}$. The quench process gives rise to the time-evolved state $|\psi(\bm{k},\tau)\rangle=e^{-i\frac{\tau}{2}\bm{h}^{f}(\bm{k})\cdot\bm{\sigma}}|\psi(\bm{k},0)\rangle$ with the flatted Hamiltonian  $\bm{h}^{f}(\bm{k})=\bm{H}^{f}(\bm{k})/|\bm{H}^{f}(\bm{k})|$ being equivalent to a rescaling of time  $t\rightarrow\tau=|\bm{H}^{f}(\bm{k})|t$, from which the Bloch vector $\bm{n}(\bm{k},\tau)=\langle\psi(\bm{k},\tau)|\bm{\sigma}|\psi(\bm{k},\tau)\rangle=-[\sin\theta(\bm{k},\tau)\cos\phi(\bm{k},\tau)$ $,\sin\theta(\bm{k},\tau)\sin\phi(\bm{k},\tau),\cos\theta(\bm{k},\tau)]$ defines a mapping $\mathcal{P}$ from a 3D torus $(k_{x},k_{y},\tau)\in T^{3}$ to the Bloch sphere $\bm{n}\in S^{2}$. For a topologically trivial $\mathcal{H}^{i}$, i.e. its Chern number $C_{i}=0$, the map $\mathcal{P}$ classified as the homotopy group  $\pi_{3}(S^{2})=\mathbb{Z}$ \cite{nakahara2018geometry} is associated with a dynamical Hopf invariant, which counts the linking number $Lk$ of the preimages of any two distinct regular values $\bm{\xi}_{1}$ and $\bm{\xi}_{2}\in S^{2}$. The Hopf invariant can be characterized by the Chern-Simons integral
\begin{equation}\label{1}
Cs=\dfrac{1}{4\pi^{2}}\int_{BZ}dk_{x}dk_{y}\int_{0}^{2\pi}d\tau A_{\mu}J_{\mu},\:(\mu=k_{x},k_{y},\tau),
\end{equation}
 here  $A_{\mu}=-i\langle\psi(\bm{k},\tau)|\partial_{\mu}|\psi(\bm{k},\tau)\rangle$ is the Berry connection and $\bm{J}=(J_{k_{x}},J_{k_{y}},J_{\tau}):\,J_{\mu}=\epsilon^{\mu\nu\rho}\partial_{\nu}A_{\rho}$ represents the Berry curvature marking the direction of preimages (Levi-Civita symbol $\epsilon^{\mu\nu\rho}$ is used). $Cs$ also mathematically proved to be the post-quench Chern number $C_{f}$ \cite{wang2017scheme}. Such a Hopf invariant has been measured recently in a Haldane-type ultracold atom system by the Hamburg group \cite{tarnowski2019measuring}. For the case of $C_{i}\neq0$, $Cs$ is not a Hopf  invariant, the mapping $\mathcal{P}$ extends to the homotopy group $\tau_{3}(S^{2})$ \cite{fox1948homotopy}  but can be restored to $\pi_{3}(S^{2})$ by clinging a complementary patch with monopole-like spin configuration to the original Brillouin zone (BZ) (seen as a gauge fixing) \cite{chen2020linking}. However, the different clinging operations will supply different preimage straight lines along the time direction, indicating a $2C_{i}$ uncertainty for such a linking number:\,\,$Lk=(C_{f}-C_{i})\mathrm{mod}(2C_{i})$.
Based on an elaborate loop-unitary trick,
 the change of the Chern number was also extracted by a 3-winding number $W_{3}=C_{f}-C_{i}$ stemming from the homotopy group $\pi_{3}[SU(2)]=\mathbb{Z}$ \cite{hu2020topological}. But in another way, it will be exciting to  establish a direct correspondence between equilibrium and nonequilibrium topological invariants without recourse to any homotopy descriptions. 

For the case of $C_{i}=0$, one or two isolated instances 
on the other hand pointed out the intimate connection between the links and the vortices in azimuthal phase profile $\phi(\bm{k}, \tau)$ by choosing the north and the south pole on the Bloch sphere as image points, but merely confined within superficial analysis \cite{yu2017phase, tarnowski2019measuring}. How far this relationship can be generalized to topological initial states has remained unclear.  Concretely, the vorticity at $(\bm{k}^{*},t^{*})$ can be calculated as
\begin{equation}\label{2}
\nu(\bm{k}^{*},t^{*})=\lim_{q\to 0}\frac{1}{2\pi}\int_{0}^{2\pi}\frac{\partial\phi(\bm{k},t^{*})}{\partial\alpha}d\alpha, 
\end{equation}
where the points nearby $\bm{k}^{*}$ are $\bm{k}^{*}+q(\cos\alpha,\sin\alpha)$.
Two kinds of vortices, i.e., the static and dynamical vortices have been captured in experiments on shaken hexagonal optical lattice \cite{flaschner2018observation, tarnowski2019measuring}.  Against such realistic background, we will pursue the relevant research further in this paper.

\begin{figure}
	\scalebox{0.5}[0.5]{\includegraphics{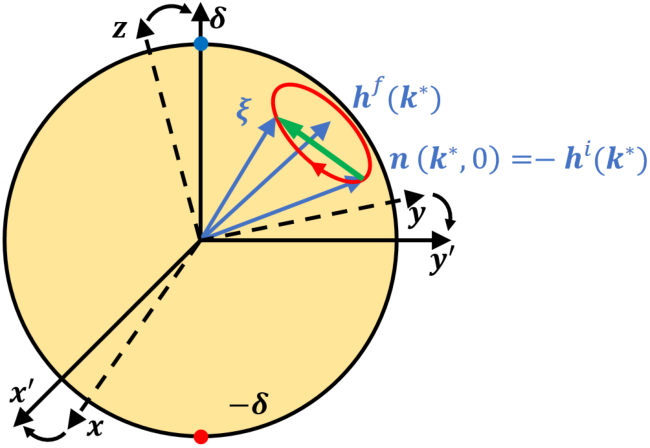}}
	\caption{\label{fig1}(Color online) Sketch of the momentum constraint. For a certain $\bm{k}^{*}$, $\bm{\xi}+\bm{h}^{i}(\bm{k}^{*})$ (the green vector) must be perpendicular to $\bm{h}^{f}(\bm{k}^{*})$ provided the Bloch vector $\bm{n}(\bm{k}^{*},0)$ reaches $\bm{\xi}$ along the red loop at any $\tau^{*}$. Besides, rotating the original spin basis ($\bm{x},\bm{y},\bm{z}$) to the new one ($\bm{x}^{\prime},\bm{y}^{\prime},\bm{\delta}$) with the redefined north pole $\bm{\delta}$ and south pole $-\bm{\delta}$ will favour the relevance between the link and the vortex.}
\end{figure}

\section{\uppercase\expandafter{\romannumeral2}.~~Orthogonal links arising from parallel quench}
First, for an abrupt quench with arbitrary $|\psi(\bm{k},0)\rangle$, $\bm{h}^{i}(\bm{k})$, and $\bm{h}^{f}(\bm{k})$,  since the Bloch vector $\bm{n}(\bm{k},\tau)$ always precesses around $\bm{h}^{f}(\bm{k})$, the preimages of any $\bm{\xi} \in S^{2}$ must meet the momentum constraint as depicted in Fig.~\ref{fig1}
\begin{equation}\label{3}
[\bm{\xi}+\bm{h}^{i}(\bm{k}^{*})]\cdot\bm{h}^{f}(\bm{k}^{*})=0.
\end{equation} 
Second, the Brouwer theorem \cite{hatcheralgebraic} for a quench process with $C_{i~\mathrm{or}~f}\neq0$ promises at least one fixed point  $\bm{k}^{s}\in\{\bm{k}\big|\bm{h}^{i}(\bm{k})\cdot\bm{h}^{f}(\bm{k})=\pm1\}$, which means the preimages $\mathcal{P}^{-1}[\bm{\xi}_{1}=\bm{n}(\bm{k}^{s},\tau^{*})]$ inevitably possess a straight line in time direction. The line combined with a preimage curve offered by any other point  $\bm{\xi}_{2}\in S^{2}$ wraping around this line makes a ``nice'' link owning the Gauss linking number (see Appendix B)
 \begin{equation}\label{4}
 Lk=\mathrm{sgn}(J_{\tau}^{s}) \omega^{J},
 \end{equation} 
here $J_{\tau}^{s}$ labels the direction of the straight line, and $\omega^{J}$ is precisely the winding number of the preimage curve in the BZ. Eq.~(\ref{3}) and (\ref{4})  together motivate us to adopt a more concrete quench protocol:
\begin{equation}\label{5}
\frac{1}{2}\bm{H}^{i}(\bm{k})\cdot\bm{\sigma}\rightarrow\frac{1}{2}\bm{H}^{f}(\bm{k})\cdot\bm{\sigma}=\frac{1}{2}[\bm{H}^{i}(\bm{k})+f(\bm{k})\bm{\delta}]\cdot\bm{\sigma}, 
\end{equation}
 where the quench quantities $f(\bm{k})\bm{\delta}$ with a same unit vector $\bm{\delta}$ are parallel or antiparallel to each other for all $\bm{k}$, thus we call it ``parallel quench'', based on which  the redefined north ($\bm{\delta}$) and south ($-\bm{\delta}$) poles in the rotated spin axes $(\bm{x}^{\prime},\bm{y}^{\prime},\bm{\delta})$ as shown in Fig.~\ref{fig1} will
  produce standard preimages. Exactly, inserting $\bm{n}(\bm{k}^{*},\tau^{*})=\pm\bm{\delta}$ into Eq.~(\ref{3}) yields
\begin{equation}\label{6} 
|\bm{H}^{i}(\bm{k}^{\ast})|+\bm{h}^{i}(\bm{k}^{\ast})\cdot\bm{\delta}f(\bm{k}^{\ast})=\mp[f(\bm{k}^{\ast})+\bm{H}^{i}(\bm{k}^{\ast})\cdot\bm{\delta}],
\end{equation}
 factored into $\bm{h}^{i}(\bm{k}^{\ast})\cdot\bm{\delta}=\mp 1$ and $|f(\bm{k}^{\ast})|=|\bm{H}^{i}(\bm{k}^{\ast})|$. Allowing for the time evolution of $\bm{n}(\bm{k},\tau)$  we further obtain the collection of preimages in $(2+1)D$ space (see Appendix C, F)
   \begin{equation}\label{7}
  \begin{aligned}
  &\left\{
  \begin{aligned}
 \bm{k}^{s}\in\{\bm{k}^{\ast}\big|\bm{h}^{i}(\bm{k}^{\ast})\cdot\bm{h}^{f}(\bm{k}^{\ast})=\pm1\},\tau^{s}\in[0,2\pi):\bm{n}=\pm\bm{\delta},\,\,\,\,\,\,\,\,\,\,\,\,\,\,\,\,\,\,\\
  \bm{k}^{d}\in\{\bm{k}^{\ast}\big||f(\bm{k}^{\ast})|=|\bm{H}^{i}(\bm{k}^{\ast})|\},\tau^{d}=\pi:\bm{n}=-\mathrm{sgn}[f(\bm{k}^{d})]\bm{\delta},
  \end{aligned}
  \right.\\
  \end{aligned}
  \end{equation}
  from which the orthogonal link can be constructed. 
  More intriguingly, 
 its linking number proved to be directly related to pre- and post-quench Chern indices (see details in Appendix):
\begin{equation}\label{8}
\begin{aligned}
Lk&=\sum\limits_{l=1}^{M}Lk_{l}\\&=-\sum\limits_{l=1}^{M}\mathrm{sgn}[f(\bm{k}_{l}^{a})]\mathrm{sgn}[f(\bm{k}_{l}^{d})]\omega^{d}_{l}\nu^{d}(\bm{k}_{l}^{d})\nu^{s}(\bm{k}_{l}^{a})\\&=-\sum\limits_{l=1}^{M}\mathrm{sgn}[f(\bm{k}_{l}^{a})]\nu^{s}(\bm{k}_{l}^{a})=C_{f}-C_{i},
\end{aligned}
\end{equation}
here we suppose the link can be divided into $M$ sublinks, corresponding to the polarity reversal before and after quenching at $M$ fixed points: $\bm{k}^{a}\in\{\bm{k}^{s}|\bm{h}^{i}(\bm{k}^{s})\cdot\bm{h}^{f}(\bm{k}^{s})=-1\}$. Because the Berry curvature contains the information about vortices, Eq.~(\ref{8}) further indicates that the dynamical topology can be extracted from the interaction between the two types of vortices in the BZ. Concretely, the preimage straight lines place the static vortices at fixed points $\bm{k}^{s}$, while the preimage loops in the $\tau=\pi$ plane trace out the dynamical vortex contours.  For the $l$th sublink $Lk_{l}$,  the winding number of the relevant preimage subloop always meets $\omega^{J}_{l}=-\mathrm{sgn}[f(\bm{k}_{l}^{d})]\omega^{d}_{l}\nu^{d}(\bm{k}_{l}^{d})=-1$ explaining the third equal sign in Eq.~(\ref{8}), 
 where the dynamical vortex with the vorticity $\nu^{d}(\bm{k}_{l}^{d})$ has to surround the $l$th static vortex with $\nu^{s}(\bm{k}_{l}^{a})$ at $\bm{k}_{l}^{a}$, and the chirality of the $l$th dynamical vortex $\omega^{d}_{l}=-\mathrm{sgn}(\partial_{k_{\varparallel}}|\bm{H}^{f}|\big|_{\bm{k}_{l}^{d}})$ labels its moving direction (see Appendix E).
 Eq.~(\ref{8}) is the main result of this work.
By the way, the quenches from a polarized initial state pointing to one pole will permit the preimage loop in the $\tau=\pi$ plane to match the critical momentum loop in dynamical quantum phase transitions (DQPT) \cite{heyl2013dynamical},  
 as only such $\bm{n}(\bm{k}^{d},0)$ at $\bm{k}^{d}$ satisfying $\bm{h}^{i}(\bm{k}^{d})\cdot\bm{h}^{f}(\bm{k}^{d})=0$ can reach another pole after half a period. 
 
For $\bm{k}$-dependent $\bm{\delta}(\bm{k})$, a series of unitary rotations $\mathcal{U}(\bm{k},\bm{\delta}^{\prime})$: \,\,$\mathcal{U}(\bm{k},\bm{\delta}^{\prime})[\bm{H}^{i}(\bm{k})+f(\bm{k})\bm{\delta}(\bm{k})]\cdot\bm{\sigma}\mathcal{U}^{\dag}(\bm{k},\bm{\delta}^{\prime})=[\bm{H}^{i\prime}(\bm{k})+f(\bm{k})\bm{\delta}^{\prime}]\cdot\bm{\sigma}=\bm{H}^{f\prime}(\bm{k})\cdot\bm{\sigma}$ will restore it to the parallel quench about $\mathcal{H}^{\prime}$ possessing the same phase boundaries as $\mathcal{H}$.  The difference between pre- and post-quench Chern numbers remains constant, as the number of antiparallel fixed points is a rotational invariant, even if $\mathcal{U}(\bm{k},\bm{\delta}^{\prime})$ may add new singularities. Actually, in this scenario the constraint $\bm{n}(\bm{k}^{*},\tau^{*})=\pm\bm{\delta}(\bm{k}^{*})$ is identical to  $\bm{n}^{\prime}(\bm{k}^{*},\tau^{*})=\pm\bm{\delta}^{\prime}$, thus offering the same preimages (certainly the same link) as the latter. Technically, the phase vortex could be depicted in a local frame $[\bm{x}(\bm{k}^{*}),\bm{y}(\bm{k}^{*}),\bm{\delta}(\bm{k}^{*})]$ at preimage points 
$\bm{k}^{*}$.
To say the least, the parallel quench protocol is enough for multiple known models, such as the Haldane model \cite{haldane1988model, sticlet2013distant}, the Qi-Wu-Zhang (QWZ) model \cite{qi2006topological}, and the Sticlet model \cite{sticlet2012geometrical}, etc, as elaborated below. 

\begin{figure}
	\centering 
	\subfigure[]{
		\begin{minipage}[t]{0.5\linewidth}
			\centering
			\includegraphics[width=4.5cm,height=2.3cm]{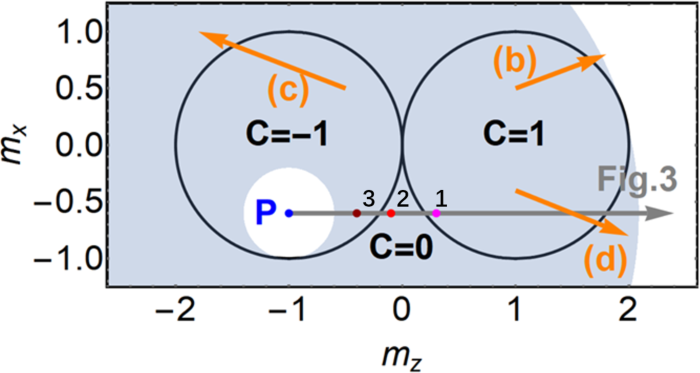}
		\end{minipage}%
	}%
	\centering 
	\subfigure[]{
		\begin{minipage}[t]{0.5\linewidth}
			\centering
			\includegraphics[width=3.8cm,height=2.5cm]{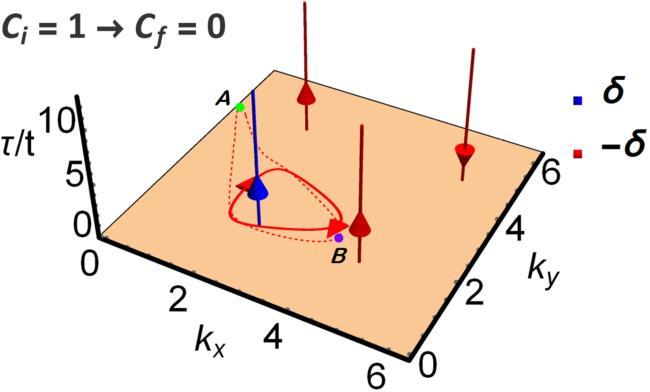}
		\end{minipage}%
	}%
	\\
	\centering 
	\subfigure[]{
		\begin{minipage}[t]{0.5\linewidth}
			\centering
			\includegraphics[width=4cm,height=3cm]{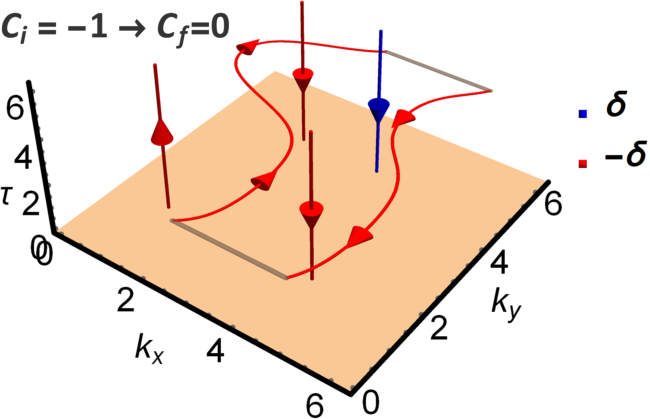}
		\end{minipage}%
	}%
	\centering 
	\subfigure[]{
		\begin{minipage}[t]{0.5\linewidth}
			\centering
			\includegraphics[width=4cm,height=3cm]{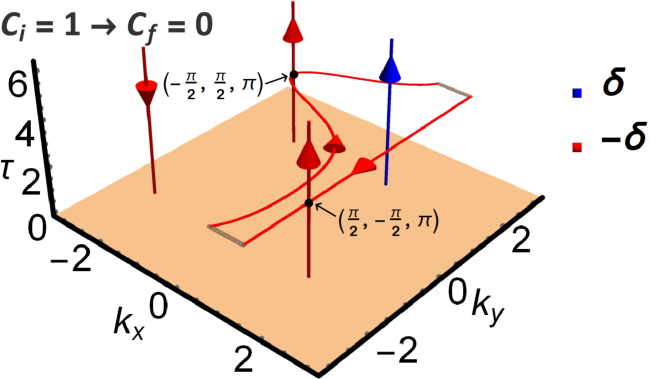}
		\end{minipage}%
	}%
\\
\centering 
\subfigure[]{
	\begin{minipage}[t]{1\linewidth}
		\flushleft
		\includegraphics[width=9cm,height=4cm]{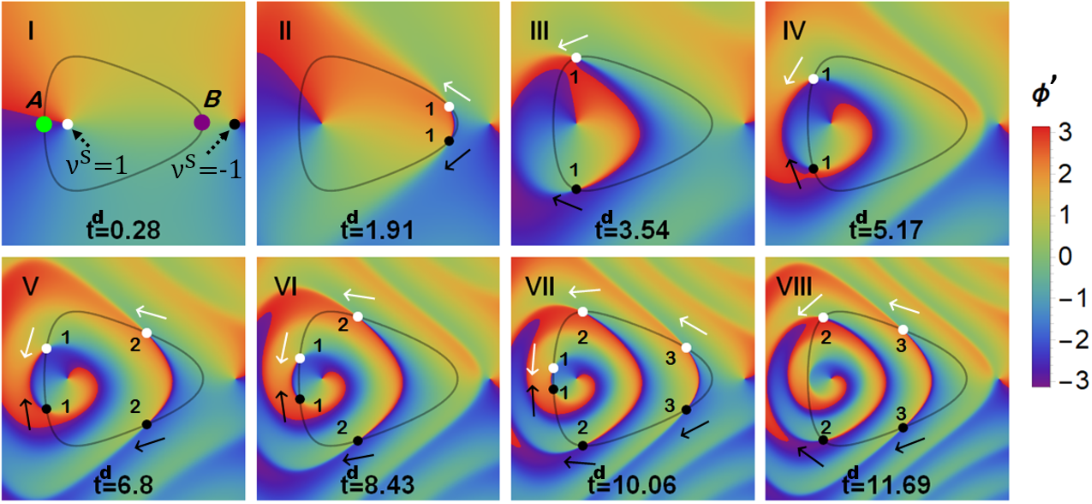}
	\end{minipage}%
}%
	\centering
	\begin{flushleft}
		\caption{\label{fig2}(Color online) (a) Phase diagram versus $m_{x}$ and $m_{z}$ for the Hamiltonian Eq.~(\ref{9}) with $t_{so}=t_{0}=1$ and $m_{y}=0$. Four quenching paths have been marked, corresponding to Fig.~\ref{fig2}(b, c, d) and Fig.~\ref{fig3}, respectively. (b, c, d) Preimages $\mathcal{P}^{-1}(\pm\bm{\delta})$ forming an orthogonal link (red curve plus blue line) with $Lk=-1$ (b), 1 (c), -1 (d): $(m^{i}_{x},m^{i}_{z},m^{f}_{x},m^{f}_{z})=(0.5,1,0.8,1.8)$ (b); $(0.5,-0.5,1,-1.8)$ (c); $(-0.4,1,-0.8,2)$ (d).
		The	dashed red curve in Fig.~\ref{fig2}(b) depicts $t^{d}=\pi/|\bm{H}^{f}(\bm{k}^{d})|$. (e) Stroboscopic observation of the azimuthal phase $\phi^{\prime}(\bm{k},t^{d})$ in the partial BZ with quench parameters as adopted in Fig.~\ref{fig2}(b). The dot A (B) in Fig.~\ref{fig2}(b, e) marks the annihilation (creation) of the dynamical vortex pair. Following the arrows nearby, The vortex pair present in different time slot travel along the gray trajectory. All the white (black) dots refer to the positive (negative) vorticity.}
	\end{flushleft}
\end{figure}

\begin{figure}
\centering 
\subfigure[]{
	\begin{minipage}[t]{0.5\linewidth}
		\centering
		\includegraphics[width=4cm,height=4cm]{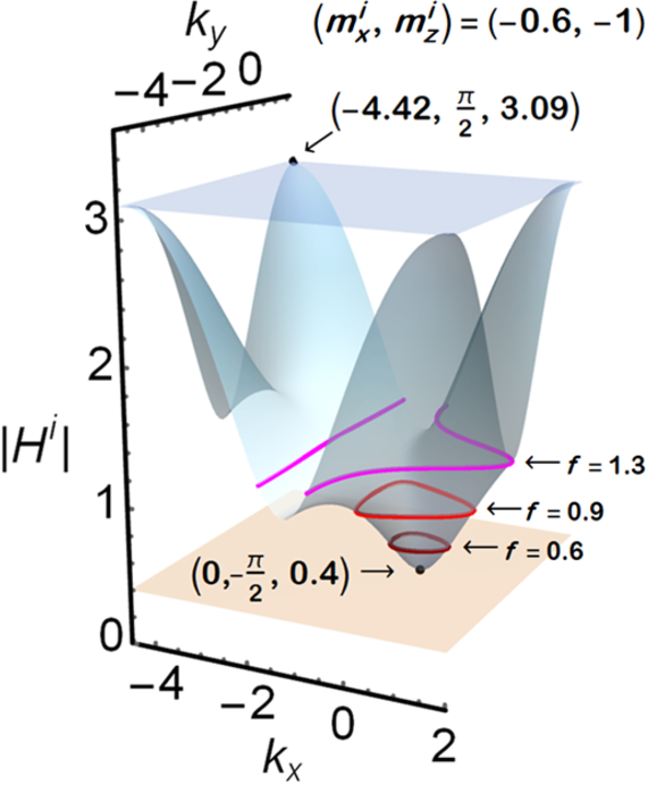}
	\end{minipage}%
}%
	\centering 
\subfigure[]{
	\begin{minipage}[t]{0.5\linewidth}
		\centering
		\includegraphics[width=4cm,height=3.2cm]{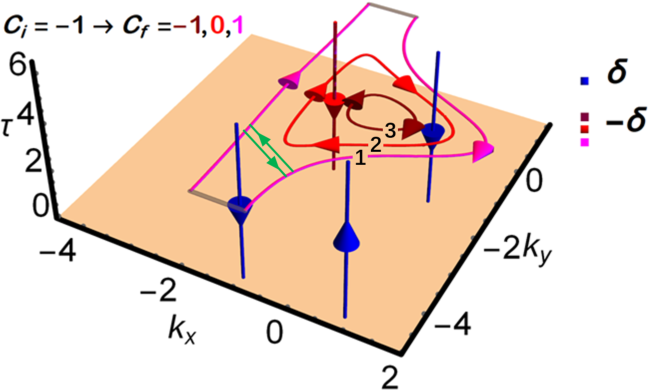}
	\end{minipage}%
}%
	\centering
	\begin{flushleft}
		\caption{\label{fig3}(Color online) (a)  Surface of $|\bm{H}^{i}|$ for $(m^{i}_{x},m^{i}_{z})=(-0.6,-1)$. 
			In order to get the preimage loop,  the quench intensity $f$ should be limited in $[|\bm{H}^{i}|_{min},|\bm{H}^{i}|_{max}]\approx[0.40,3.09]$. $(k_{x},k_{y})=(0,-\pi/2)$ and $(-4.42,\pi/2)$ denote two collapse points of the preimage loop. (b) Preimages  $\mathcal{P}^{-1}(\pm\bm{\delta})$.  $f$=0.6 (dark red), 0.9 (red) and 1.3 (magenta) give the right linking number $Lk$=0, 1 and 2, respectively. Two blue lines with arrows pointing down are encircled by a magenta curve, corresponding to a biskyrmion. Its sublinks with $Lk_{1}=Lk_{2}=1$ can be made by adding a pair of auxiliary (green) lines. The trivial (dark red)  contour signals the unchanged Chern number.}
	\end{flushleft}
\end{figure}

\section{\uppercase\expandafter{\romannumeral3}.~~Quench in variable direction}
As a benchmark, we invoke the extended QWZ model
\begin{equation}\label{9}
\begin{aligned}
&H_{x}=m_{x}+t_{so}\cos(k_{x}),\,\,H_{y}=m_{y}+t_{so}\cos(k_{y}),\\
&H_{z}=m_{z}-t_{0}[\sin(k_{x})+\sin(k_{y})]
\end{aligned}
\end{equation}
realized by the USTC-PKU group on a square Raman lattice recently \cite{yi2019observing}, where $t_{0}$ ($t_{so}$) denotes
 the spin-conserved (flip) hopping coefficient. Suddenly modulating the Zeeman constants $m_{x,y,z}$ can implement the quenches along arbitrary spin axes. To elucidate the main points, we manipulate $(m_{x},m_{z})$, but always take $t_{so}=t_{0}=1$ and $m_{y}=0$. In this case the four Dirac points $[\arccos(-m_{x}), \pm\frac{\pi}{2}]$ and $[-\arccos(-m_{x}), \pm\frac{\pi}{2}]$ correspond to four phase boundaries $m_{z}-\sqrt{1-m_{x}^{2}}\mp1=0$ and $m_{z}+\sqrt{1-m_{x}^{2}}\mp1=0$, respectively. The quench intensity $f(\bm{k})=\sqrt{(m_{x}^{f}-m_{x}^{i})^{2}+(m_{z}^{f}-m_{z}^{i})^{2}}$ ($\equiv f$) and direction $ \bm{\delta}=(\frac{m_{x}^{f}-m_{x}^{i}}{f},0,\frac{m_{z}^{f}-m_{z}^{i}}{f})$ are independent of $\bm{k}$  under every quench process.

  To form a notable contrast with previous results \cite{wang2017scheme,tarnowski2019measuring, chen2020linking,yu2017phase}, we restrict ourselves to the case of $C_{i}\neq0$, and plot the preimages $\mathcal{P}^{-1}(\pm\bm{\delta})$ in Fig.~\ref{fig2}(b, c, d) along different quenching paths as labeled in Fig.~\ref{fig2}(a). In the rotated spin axes $(\bm{x}^{\prime},\bm{y}^{\prime},\bm{\delta})$, the new north and south poles $\pm\bm{\delta}$ (see Fig.~\ref{fig1}) will be of benefit to redefine the new vortices for the azimuthal phase $\phi^{\prime}(\bm{k},t)$. Using the right-hand rule, all the orthogonal links as well as the associated vortex animations present the correct linking numbers indeed. For example, in Fig.~\ref{fig2}(e) we can easily check that $\mathrm{sgn}[f(\bm{k}^{a})]=\mathrm{sgn}[f(\bm{k}^{d})]=\nu^{s}(\bm{k}^{a})=1$, and the dynamical vortex (anti-vortex) moves around the static vortex counterclockwise (clockwise) meeting $\omega^{d}\nu^{d}(\bm{k}^{d})=1$ (so $\omega^{J}=-1$), in agreement with Eq.~(\ref{8}). The paired auxiliary lines with the opposite direction [the gray lines in Fig.~\ref{fig2}(c,d)] in the BZ can be added without restriction.  It should be noted that the fixed points determined by
  \begin{equation}\label{10}
  \left\{
  \begin{array}{cccc}
  \begin{split}
  &A\cos{k_{x}}+B(\sin{k_{x}}+\sin{k_{y}})+C=0, \\
  &\cos{k_{y}}=0, \\
  \end{split}
  \end{array}
  \right.\\
  \end{equation}\label{12}
with $A=m^{f}_{z}-m^{i}_{z}$, $B=m^{f}_{x}-m^{i}_{x}$, and $C=m^{i}_{x}m^{f}_{z}-m^{i}_{z}m^{f}_{x}$, locating the static vortices (i.e. the vertical preimages), are no longer the Dirac points unless $m_{x}^{i}=m_{x}^{f}$. The self-intersection of preimages
  $\mathcal{P}^{-1}(-\bm{\delta})$ occurs at two points $\bm{k}^{1}=(-\frac{\pi}{2},\frac{\pi}{2})$ and $\bm{k}^{2}=(\frac{\pi}{2},-\frac{\pi}{2})$ of the $\tau=\pi$ plane as shown in Fig.~\ref{fig2}(d) with zero Berry curvature $\bm{J}=0$, corresponding to an instantaneous disappearance of the static vortex at time $t_{1 or 2}=\pi/|\bm{H}^{f}(\bm{k}^{1 or 2})|$ (as shown in Appendix E). For the case of Fig.~\ref{fig2}(b) [the multiple local maxima (minima) of $|\bm{H}^{f}(\bm{k}^{d})|$ may exist in other cases], 
   $\tau^{d}=\pi$ in Eq.~(\ref{7}) implies that the time slot the $\mathcal{M}$th pair of dynamical vortices appears is  
\begin{equation}\label{11}
t^{d}\in\Big[\frac{(2\mathcal{M}-1)\pi}{|\bm{H}^{f}(\bm{k}^{d})|_{max}}, \frac{(2\mathcal{M}-1)\pi}{|\bm{H}^{f}(\bm{k}^{d})|_{min}}\Big], \mathcal{M}=1,2,3,\ldots
\end{equation}
but different time slots can overlap as shown in Fig.~\ref{fig2}(e).

Besides, the quench intensity that supports the appearance of the preimage loop (i.e., the trajectories of dynamical vortices) need to meet $\label{r} |\bm{H}^{i}(\bm{k})|_{max}>f>|\bm{H}^{i}(\bm{k})|_{min}$, which determines the post-quench parameter region provided the initial parameters have been given, e.g., the light blue region in Fig.~\ref{fig2}(a) for the starting point $(m_{x}^{i},m_{z}^{i})=(-0.6,-1)$ (the blue dot P). Fig.~\ref{fig3} exhibits that the preimage loop, if present, always make the right linking number, meanwhile forms a skyrmion spin texture in the $\tau=\pi$ plane \cite{skyrme1958non, manton2004topological}, but the preimage loop will be contracted to a collapse point gradually when $f$ approaches the extreme value. In other words, the vanishing link does not necessarily mean $C_{i}=C_{f}$.  

\begin{figure}
	\centering 
	\subfigure[]{
		\begin{minipage}[t]{0.5\linewidth}
			\centering
			\includegraphics[width=3.5cm,height=3.5cm]{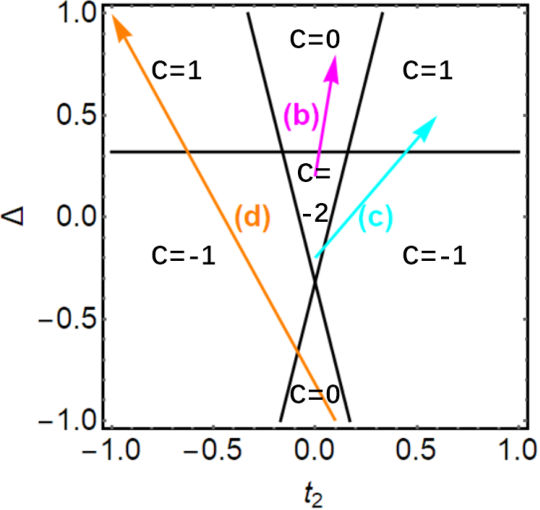}
		\end{minipage}%
	}%
	\centering 
	\subfigure[]{
		\begin{minipage}[t]{0.5\linewidth}
			\centering
			\includegraphics[width=3.8cm,height=2.9cm]{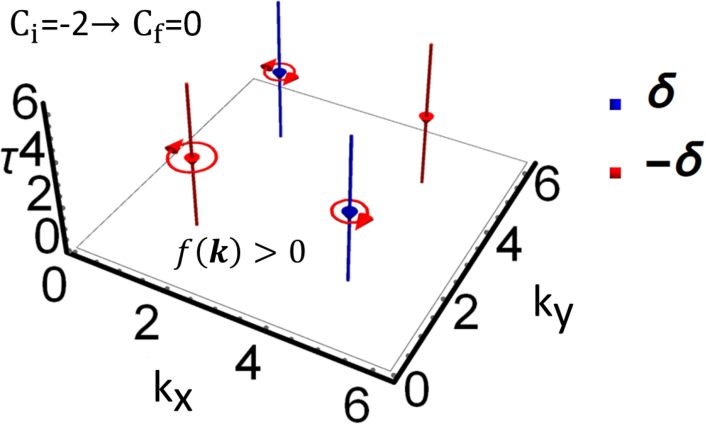}
		\end{minipage}%
	}%
	\\
	\centering 
	\subfigure[]{
		\begin{minipage}[t]{0.5\linewidth}
			\centering
			\includegraphics[width=3.8cm,height=2.9cm]{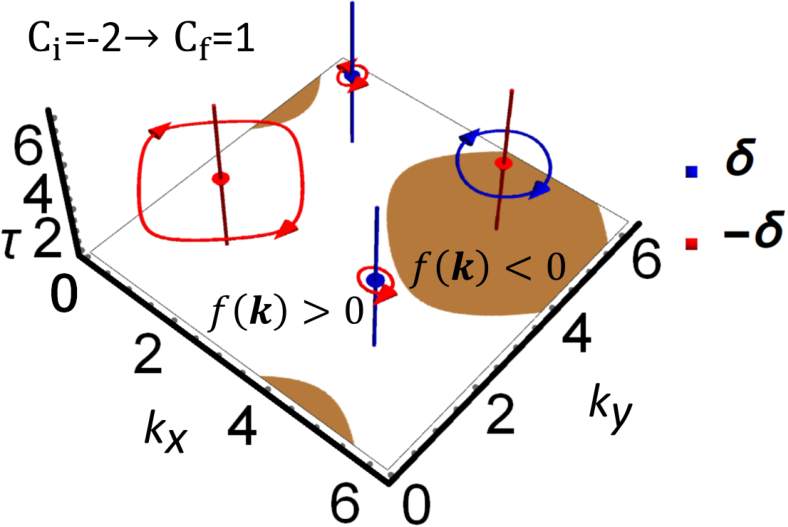}
		\end{minipage}%
	}%
	\centering 
	\subfigure[]{
		\begin{minipage}[t]{0.5\linewidth}
			\flushleft
			\includegraphics[width=4cm,height=3cm]{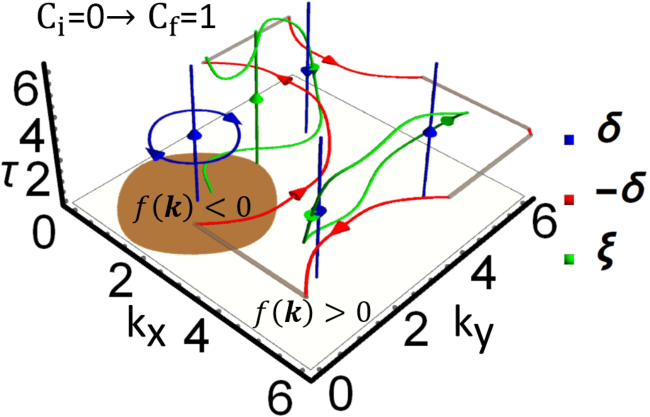}
		\end{minipage}%
	}%
	\centering
	\begin{flushleft}
		\caption{\label{fig4}(Color online) (a) Phase diagram of the Sticlet model with $t_{1}=1$ and $t_{3}=-0.08$. $(\Delta^{i},t_{2}^{i},\Delta^{f},t_{2}^{f})=(0.2,0,0.8,0.1)$ (b), $(-0.2,0,0.5,0.6)$ (c), $(-1,0.1,1,-1)$ (d). (b) Separate sublinks with $Lk_{1}=Lk_{2}=1$ (exclude the self-winding of red lines). (c) Separate sublinks with $Lk_{1}=Lk_{2}=Lk_{3}=1$. The preimage loop in the region  $f(\bm{k})<0$ ($>0$) belongs to the north (south) pole. (d) Preimages of $\bm{\delta}$, $-\bm{\delta}$, and $\bm{\xi}=(0,0.92,0.39)$. $\bm{\xi}$ offers a straight line located on the fixed loop $f(\bm{k}^{\mathcal{A}})=0$. Any two of the three points creat a link with $Lk=1$ (the paired auxiliary lines are omitted).}
	\end{flushleft}
\end{figure}

\section{\uppercase\expandafter{\romannumeral4}.~~Varying quench intensity and higher Chern index}
The second evidence is the Sticlet model realized on a triangular lattice with two orbitals on each site \cite{sticlet2012geometrical}
\begin{equation}\label{12}
\begin{aligned}
H_{x}=&2t_{1}\cos(k_{x}),\,\,\,\,\,\,\,H_{y}=2t_{1}\cos(k_{y}),\\
H_{z}=&\Delta+2t_{2}[\sin(k_{x})+\sin(k_{y})]\\&+2t_{3}[\cos(k_{x}+k_{y})-\cos(k_{x}-k_{y})],
\end{aligned}
\end{equation}
which degenerates to the QWZ model when $t_{3}=0$. In this model the four fixed points coincide with the Dirac points $(k_{x},k_{y})=(\pm\frac{\pi}{2},\pm\frac{\pi}{2})$, and the phase boundaries $\bigtriangleup=-4 t_{3}$ and $\bigtriangleup=4(t_{3}\pm t_{2})$  are determined by gapless condition. The quenches on the phase diagram Fig.~\ref{fig4}(a) 
	suggest $\bm{\delta}=\bm{z}$ and  $f(\bm{k})=\Delta^{f}-\Delta^{i}+2(t_{2}^{f}-t_{2}^{i})[\sin({k_{x}})+\sin({k_{y}})]$. The relation in Eq.~(\ref{7}) enforces the preimage loops inside the region $f(\bm{k})<0$ ($>0$) to be attached to the north (south) pole. No doubt, the orthogonal link in Fig.~\ref{fig4}(b, c, d) renders a  linking number correctly, and the dynamic topology will continue to be entirely encoded in the $2D$ animation about vortices (here we omitted the $2D$ vortex pictures). It is obvious that the varying quench intensity may yield an accidental fixed loop with $f(\bm{k}^{\mathcal{A}})=0$, and each orthogonal sublink can not span this border line [see Eq.(A24)]. 
For any $\bm{\xi}\in S^{2}$, if there is a $\bm{k}^{\mathcal{A}\prime}\in\{\bm{k}^{\mathcal{A}}|f(\bm{k}^{\mathcal{A}})=0\}$ meeting $\bm{\xi}=\bm{n}(\bm{k}^{\mathcal{A}\prime},\tau^{\ast})$ \cite{[In the case of $C_{i}=0\mathrm{,}$ assume that
$\bm{\xi}=\bm{n}(\bm{k}^{\mathcal{A}}\mathrm{,}\tau^{\ast})$ holds for all $\bm{k}^{\mathcal{A}}$ on the whole fixed loop$\mathrm{,}$ $\mathcal{P}^{-1}(\bm{\xi})$ will supply a preimage torus $T^{2}$. As $\partial_{\tau}\bm{n}(\bm{k}\mathrm{,}\tau)|_{q}=\partial_{k_{\varparallel}}\bm{n}(\bm{k}\mathrm{,}\tau)|_{q}=0$ for any point $q\in T^{2}\mathrm{,}$ $\bm{\xi}$ is no longer a regular value in the terminology of differential topology$\mathrm{,}$ and the Berry curvature $\bm{J}(q)=0$ makes the link involving $\bm{\xi}$ become an illness$\mathrm{,}$ although the Chern-Simons integral $C_{s}$ can still give a Hopf invariant in this case] AAA}, the self-intersection of preimages $\mathcal{P}^{-1}(\bm{\xi})$ will occur again, because an extra preimage straight line offered by $\bm{n}(\bm{k}^{\mathcal{A}\prime},\tau^{\ast})$ joins in [see Fig.~\ref{fig4}(d)], contributes to a right linking invariant. Notably, rewriting the quench term as $f(\bm{k})=\zeta[\epsilon+2(\sin({k_{x}})+\sin({k_{y}}))]$,   when $\zeta\rightarrow\infty$, the momentum constraint $[\pm\bm{z}+\bm{h}^{i}(\bm{k}^{*})]\cdot\bm{h}^{f}(\bm{k}^{*})=0$ from Eq.~(\ref{3}) can be expanded to the lowest order of $1/\zeta$:\,\,$\mathrm{sgn}[\epsilon+2(\sin({k_{x}^{*}})+\sin({k_{y}^{*}}))][h^{i}_{\bm{z}}(\bm{k}^{*})\pm 1]+\mathcal{O}(1/\zeta)=0$, which implies that, under any extreme quenches from a nonpolarized  initial state, the preimages of the two poles will converge to the fixed points and the fixed loop if the latter exists.  In contrast,  the critical momentum loop in DQPT will converge to the fixed loop and another loop with $h^{i}_{\bm{z}}(\bm{k})=0$ in terms of $\bm{h}^{i}(\bm{k})\cdot\bm{h}^{f}(\bm{k})=\mathrm{sgn}[\epsilon+2(\sin(k_{x})+\sin(k_{y}))]h^{i}_{\bm{z}}(\bm{k})+\mathcal{O}(1/\zeta)$.

\section{\uppercase\expandafter{\romannumeral5}.~~Experimental feasibility and conclusion}
The dilute ultracold atomic gases in optical lattices  provide versatile platforms  for exploring the $2D$ topological quantum matters out of equilibrium \cite{jotzu2014experimental, wu2016realization, cooper2019topological, eckardt2017colloquium, gross2017quantum, schafer2020tools}. The optical Raman lattice scheme will facilitate the flexible choice of the quench direction $\bm{\delta}$ by quickly varying the two-photon detuning and phases of Raman couplings together \cite{yi2019observing,zhang2019dynamical}. Applicating a  coherent Raman pulse \cite{yi2023extracting},  a momentum and time-resolved state tomography is expected to map out  the preimage contours in $T^{3}$ even for $\bm{k}$-dependent $\bm{\delta}(\bm{k})$ \cite{hauke2014tomography,flaschner2016experimental, yi2019observation}. The adiabatic preparation of topological phases with $C_{i}\neq0$ can be implemented by bringing in a conjugated duplicate with $-C_{i}$ \cite{barbarino2020preparing}. Therefore, the extraction of the dynamical topology under the parallel quench protocol will be achieved experimentally via state-of-the-art techniques.

In summary, we present a practical procedure for getting a rather intuitive physical picture of generalized dynamical Hopf insulators, although its Chern-Simons integral $Cs$ may not be a topological invariant \cite{chen2020linking}. The orthogonal links in $(2+1)D$ space provide various skyrmion-like configurations in $\tau=\pi$ place,  further a vivid $2D$ visualization about phase vortices. The fixed point plays a critical role so that the associated preimages explicitly defined the linking number in this parallel quench protocol, whose upgraded version will cover all the sudden quenching patterns with $\mathcal{H}^{f}=\mathcal{H}^{i}+\frac{1}{2}f(\bm{k})\bm{\delta}(\bm{k})\cdot\bm{\sigma}$ (may be extend to the stroboscopic analogues in floquet topological insulators, wherein $\mathcal{H}^{f}$ refers to an effective Hamiltonian).  Our findings perfect the urgent research of the $(2+1)D$ dynamical topology, and are expected to open up a new route to characterize topological invariants via the direct development of the space-time link in experimental nonequilibrium settings.

 \section{Acknowledgments} 
This work was supported by the National Science Foundation of China (Grant No. 11664021, 61565008, 11365013), by the Opening Foundation of State Key Laboratory of Surface Physics (Grant No. KF2017\underline{ }06), and by Yunnan Ten Thousand Talents Plan Young and Elite Talents Project (Grant Number:YNWR-QNBJ-2018-121).

\appendix
\setcounter{table}{0}  
\setcounter{figure}{0}
\setcounter{equation}{0}

\renewcommand{\thetable}{A\arabic{table}}
\renewcommand{\thefigure}{A\arabic{figure}}
\renewcommand{\theequation}{A\arabic{equation}}
\begin{widetext}
\section{Appendix A:\,\,\,Tangent vector of preimages $(J_{k_{x}},J_{k_{y}},J_{\tau})$}
First, we briefly explain why the Berry curvature $\bm{J}=(J_{k_{x}},J_{k_{y}},J_{\tau})$: 
\begin{equation}\label{a1}
	J_{\mu}=\frac{1}{4}\epsilon^{\mu\nu\rho}\bm{n}\cdot(\partial_{\nu}\bm{n}\times\partial_{\rho}\bm{n}),\,\,\,\,\,\,(\mu=k_{x},k_{y},\tau)
\end{equation}
is indeed the tangent vector of preimage curves. For ease of notation, we will often use the shorthand in absence of ambiguity, say, $J_{\mu}\equiv J_{\mu}(\bm{k},\tau)$ hereafter. Consider a constant vector $\bm{a}=(a_{x},a_{y},a_{z})$ on the Bloch sphere. The preimages provided by the constraint $\bm{n}(\bm{k}^{*},\tau^{*})=\bm{a}$ can be regarded as the intersection between the two surfaces $n_{x}(\bm{k},\tau)=a_{x}$ and $n_{y}(\bm{k},\tau)=a_{y}$, of which the normal vectors are $\bm{b}_{x}=\partial_{k_{x}}n_{x}\bm{e}_{x}+\partial_{k_{y}}n_{x}\bm{e}_{y}+\partial_{\tau}n_{x}\bm{e}_{\tau}$ and $\bm{b}_{y}=\partial_{k_{x}}n_{y}\bm{e}_{x}+\partial_{k_{y}}n_{y}\bm{e}_{y}+\partial_{\tau}n_{y}\bm{e}_{\tau}$, respectively. Further more, the tangent vector $\bm{l}_{xy}$ of the preimage curve  $\bm{n}(\bm{k}^{*},\tau^{*})=\bm{a}$ at any point $(\bm{k}^{*},\tau^{*})$ can be expressed as
\begin{equation}\label{a2}
	\begin{aligned}
		&\bm{l}_{xy}(\bm{k}^{*},\tau^{*})=\bm{b}_{x}(\bm{k}^{*},\tau^{*})\times\bm{b}_{y}(\bm{k}^{*},\tau^{*})=[(\partial_{k_{y}}n_{x}\partial_{\tau}n_{y}-\partial_{\tau}n_{x}\partial_{k_{y}}n_{y})\bm{e}_{x}\\
		&+(\partial_{\tau}n_{x}\partial_{k_{x}}n_{y}-\partial_{k_{x}}n_{x}\partial_{\tau}n_{y})\bm{e}_{y}+(\partial_{k_{x}}n_{x}\partial_{k_{y}}n_{y}-\partial_{k_{y}}n_{x}\partial_{k_{x}}n_{y})\bm{e}_{\tau}]|_{(\bm{k}^{*},\tau^{*})}.
	\end{aligned}
\end{equation}
In the same way, with the help of the two surfaces $n_{y}(\bm{k},\tau)=a_{y}$ and $n_{z}(\bm{k},\tau)=a_{z}$ [or $n_{z}(\bm{k},\tau)=a_{z}$ and $n_{x}(\bm{k},\tau)=a_{x}$], we get this tangent vector  $\bm{l}_{y\tau}$ (or $\bm{l}_{\tau x}$). $\bm{l}_{xy}$, $\bm{l}_{y\tau}$ and $\bm{l}_{\tau x}$ are in the same direction at the point $(\bm{k}^{*},\tau^{*})$, hence the Berry curvature  $\bm{J}=(J_{k_{x}},J_{k_{y}},J_{\tau})=\frac{1}{4}(n_{x}\bm{l}_{y\tau}+n_{y}\bm{l}_{\tau x}+n_{z}\bm{l}_{xy})$ is also the tangent vector of the preimage curve, but its norm $|\bm{J}|$ does not have to equal to $1$.

\section{Appendix B:\,\,\,From Gauss linking number to winding number}

The Gauss linking number of two disjoint oriented curves $\gamma_{a}$ and $\gamma_{b}$ is defined as a double integral \cite{gauss2013werke}: 
\begin{equation}\label{b1}
	Lk=\frac{1}{4\pi}\int_{\gamma_{a}}\int_{\gamma_{b}}(d\bm{r}_{a}\times d\bm{r}_{b})\cdot\frac{\bm{r}_{a}-\bm{r}_{b}}{|\bm{r}_{a}-\bm{r}_{b}|^{3}},
\end{equation}
\begin{figure}
	\centering 
	\subfigure[]{
		\begin{minipage}[t]{0.3\linewidth}
			\centering
			\includegraphics[width=5.4cm,height=4.2cm]{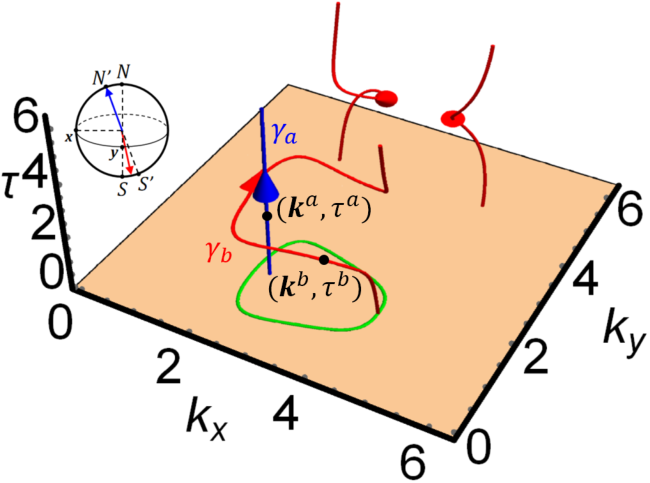}
		\end{minipage}%
	}%
	\centering 
	\subfigure[]{
		\begin{minipage}[t]{0.3\linewidth}
			\centering
			\includegraphics[width=5.4cm,height=4.2cm]{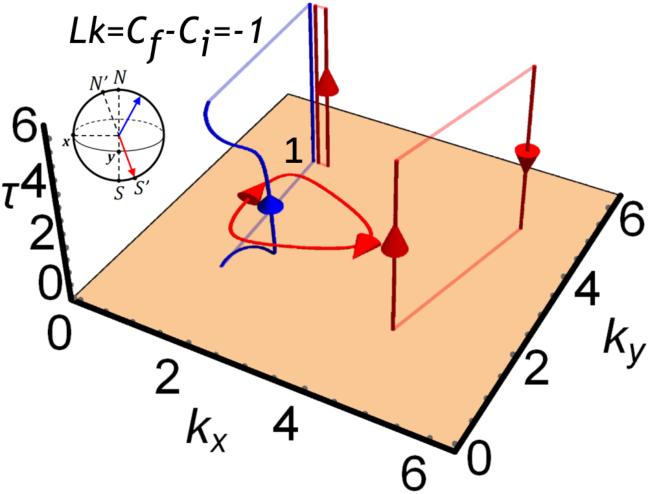}
		\end{minipage}%
	}%
	\centering
	\subfigure[]{
		\begin{minipage}[t]{0.3\linewidth}
			\centering
			\includegraphics[width=5.4cm,height=4.2cm]{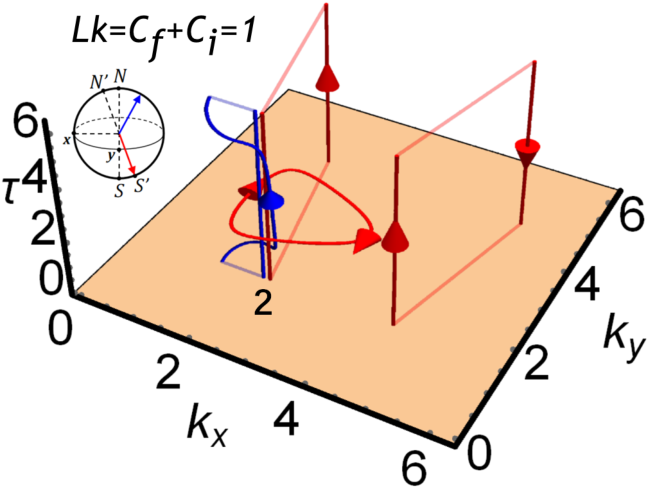}
		\end{minipage}%
	}%
	\centering
	\begin{flushleft}
		\caption{\label{figs1}(Color online) (a) Preimages of two points on the Bloch sphere with the original coordinate parameters $(\theta,\phi)=(2.78, \pi)$ (the blue line) and $(0.14, 0)$ (the red curves). The former denotes the north pole $N^{\prime}$ in the new coordinate frame after rotation. The green loop is the projection of the partial red curve $\gamma_{b}$ on the $\tau=0$ plane, which wraps around the antiparallel fixed point $\bm{k}^{a}$ once. (b, c) As a contrast, we choose two other points: one point $(\theta,\phi)=(0.36, 0)$ is the new south pole $S^{\prime}$ offering the red preimages; another point  $(\theta,\phi)=(2.64,0)$ offers the blue preimages. The periodicity of the torus  $T^{3}$ accounts for the paired auxiliary lines in lighter colors.  Adding the vertical auxiliary lines at the fixed point $1$ and $2$ gives rise to the link with $Lk=-1$ (b) and $1$ (c), respectively. The latter will become an illness when the blue image point approachs $N^{\prime}$. For these plots, the quench parameters $(m^{i}_{x},m^{i}_{z},m^{f}_{x},m^{f}_{z})=(0.5,1,0.8,1.8)$ are taken from  the Hamiltonian Eq.~(9).}
	\end{flushleft}
\end{figure}
where $\bm{r}_{a}$ and $\bm{r}_{b}$ denote the points on the two curves $\gamma_{a}$ and $\gamma_{b}$, respectively. In the case of finite open curves the Gauss linking number is not a topological invariant, and, generally, not an integer. Over the past decade, this classic concept has been extended from $R^{3}$ to $T^{3}$  \cite{panagiotou2015linking, panagiotou2018linking}, just like the objects we deal with.
Under arbitrary sudden quenches, for any point $\bm{\xi}$ on the Bloch sphere,  the vertical projection of the preimage $\mathcal{P}^{-1}[\bm{\xi}=\bm{n}(\bm{k}^{*},\tau^{*})]$ on the $\bm{k}$ space should satisfy Eq.(3) of the main text. This does not mean that the link is easy to observe and handle, as the preimages with different $\tau^{*}$ in the 3D $(\bm{k},\tau)$ space may still look disorganized. A prospective strategy is to pick out a particular $\bm{\xi}_{a}$ that corresponds to fixed points, which must exist under the case of $C_{i~\mathrm{or}~f}\neq0$, such that this point plus any other point $\bm{\xi}_{b}$ on the Bloch sphere will produce a more regular linking pattern. To gain intuition, here we show that, if a link is consisted of a vertical line at the fixed point and an open curve winding around this line, its Gauss linking number can be degenerated to a winding number of the vertical projection of the curve and is an integer even if no vertical auxiliary line is added \cite{chen2020linking}. As illustrated in Fig.\ref{figs1}(a) wherein $\bm{r}_{a}=(k_{x}^{a},k_{y}^{a},\tau^{a})$ and $\bm{r}_{b}=(k_{x}^{b},k_{y}^{b},\tau^{b})$, it is sensible to extend the upper and lower bounds of integral on the vertical line $\gamma_{a}$ to the infinity in view of the periodicity of $\tau$ \cite{ahmad2020characterization}. In Fig.\ref{figs1}(a) the two endpoints of the partial red curve  $\gamma_{b}$  winding around $\gamma_{a}$ possess the same momentum coordinate, 
thus
\begin{equation}\label{b2}
	\begin{aligned}
		Lk&=\frac{\mathrm{sgn}(J_{\tau}^{a})}{4\pi}\int_{\gamma_{b}}\int_{-\infty}^{+\infty}\frac{M_{x}dk_{y}-M_{y}dk_{x}}{[M_{x}^{2}+M_{y}^{2}+(\tau^{b}-\tau^{a})^{2}]^{\frac{3}{2}}}d\tau^{a}\\
		&=\mathrm{sgn}(J_{\tau}^{a})\cdot \omega^{J},
	\end{aligned}
\end{equation}
where $J_{\tau}^{a}$ [ $\mathrm{sgn}(X)$ denotes $X$'s sign] is the Berry curvature on $\gamma_{a}$ (see Appendix E for details), and $\omega^{J}=\frac{1}{2\pi}\ointctrclockwise\frac{M_{x}dk_{y}-M_{y}dk_{x}}{M_{x}^{2}+M_{y}^{2}}$ expresses the winding number of the 2D vector $(M_{x},M_{y})=(k_{x}^{b}-k_{x}^{a},k_{y}^{b}-k_{y}^{a})$, which counts how many times  the projection of the curve $\gamma_{b}$  wraps around the point $\bm{k}^{a}$ in the $\tau=0$ plane. However, if $\bm{\xi}_{a}$, such as $S^{\prime}$, inversely maps into the inappropriate fixed points, the preimage straight lines located at these fixed points will not be enclosed by the preimage curves offered by another image point $\bm{\xi}_{b}$ as shown in Fig.\ref{figs1}(b, c), where the vertical auxiliary lines added at the fixed point $1$ ($2$) was used to help construct a 
link with linking number $Lk=C_{f}-C_{i}=-1$ ($Lk=C_{f}+C_{i}=1$) \cite{chen2020linking}. The distinction between the two scenarios is that, when the blue image point on the Bloch sphere approaches $N^{\prime}$, the blue preimage curve will gradually overlap with the vertical auxiliary lines in Fig.\ref{figs1}(c) so that the link with $Lk=1$ will become ill-defined, while the link with $Lk=-1$  in Fig.A1(b) still behaves well. Removing the auxiliary lines from Fig.\ref{figs1}(b, c), it is found that Eq.(\ref{b1}) is unable to offer a winding number naturally without topological deformation. We will prove below that $Lk=C_{f}- C_{i}$ is universally valid when choosing the two new poles as image points.

\section{Appendix C:\,\,\,Looking for the standard preimages}
Inspired by the above analyses, we expect that the linking structure can be more tractable. Actually, if the quench is limited to the parallel quench in the
rotated spin axes $(\bm{x'},\bm{y'},\bm{\delta})$:
\begin{equation}\label{c1} 
	\bm{h}^{i}= \bm{H}^{i}/|\bm{H}^{i}|=(H_{x'},H_{y'},H_{\delta}^{i})/|\bm{H}^{i}| \rightarrow \bm{h}^{f}= \bm{H}^{f}/|\bm{H}^{f}|=(H_{x'},H_{y'},H_{\delta}^{f})/|\bm{H}^{f}|
\end{equation}  
with
\begin{equation} \label{c2} 
	H_{\delta}^{f}=H_{\delta}^{i}+f(\bm{k}),
\end{equation}
$\bm{h}^{i}(\bm{k})$, $\bm{h}^{f}(\bm{k})$ and $\pm\bm{\delta}$ are coplanar for all $\bm{k}$. Because the time- and  quasimomentum-dependent Bloch vectors in an orthogonal basis can be given by \cite{ezawa2018topological,yang2018dynamical} 
\begin{equation}\label{c3}
	\bm{n}(\bm{k},\tau)=\bm{e}_{1}+\bm{e}_{2}\cos\tau+\bm{e}_{3}\sin\tau,\,\,\,~\mathrm{here}~ \bm{e}_{1}=-\bm{h}^{f}(\bm{h}^{i}\cdot\bm{h}^{f}),\bm{e}_{2}=\bm{h}^{f}(\bm{h}^{i}\cdot\bm{h}^{f})-\bm{h}^{i},\bm{e}_{3}=\bm{h}^{f}\times\bm{h}^{i},
\end{equation}
the constraint $\bm{n}(\bm{k}^{*},\tau^{*})=\pm\bm{\delta}$ means that the $\bm{e}_{3}$ component must be equal to zero, i.e., $\tau=m\pi$ ($\bm{h}^{f}\times\bm{h}^{i}=0$ corresponds to the vertical preimages).  Combined with Eq.(3) in the main text, $m$
can only be a positive odd number and the preimage loops in the $\tau=\pi$ plane exactly belong to $-\mathrm{sgn}[f(\bm{k}^{*})]\bm{\delta}$. 
The conclusion of this section is that, in the case of parallel quenches, if we set $\pm\bm{\delta}$ as the new north and south poles, their preimages except the vertical lines will totally fall into the $\tau=\pi$ plane given that $\tau\in[0,2\pi)$, so the link must be orthogonal. The preimage loops in the $\tau=\pi$ plane will be closed by just adding paired auxiliary lines at the boundary of the Brillouin zone, then the Gauss linking number is not only an integer, but also a topological invariant.

\section{Appendix D:\,\,\,Vorticity as a topological charge}

The orthogonal link marks the trajectories of phase vortices including the static vortices at the fixed points and the dynamical vortices at the other $\bm{k}^{*}$ points.  Although the vortex can be defined in any coordinate frame, rotating the axes will directly  correlate it to the link. Here we will argue that the vorticity in the azimuthal phase profile only takes on quantized values $0$ and $\pm1$. Note that the Bloch vector can be expressed as $\bm{n}(\bm{k},t)=n_{x'}(\bm{k},t)\bm{x'}+n_{y'}(\bm{k},t)\bm{y'}+n_{\delta}(\bm{k},t)\bm{\delta}$ at time $t$ in the rotated basis $(\bm{x'},\bm{y'},\bm{\delta})$, where $n_{\eta}(\bm{k},t)=\langle\psi(\bm{k},t=0)|e^{i\frac{t}{2}\bm{H}^{f}(\bm{k})\cdot\bm{\sigma}}\bm{\eta}\cdot\bm{\sigma}e^{-i\frac{t}{2}\bm{H}^{f}(\bm{k})\cdot\bm{\sigma}}|\psi(\bm{k},t=0)\rangle$ $(\bm{\eta}=\bm{x'},\bm{y'},\bm{\delta})$. 
Focusing on a special point $(\bm{k}^{*},t^{*})$ that satisfies $\bm{n}(\bm{k}^{*},t^{*}\equiv \tau^{*}/|\bm{H}^{f}(\bm{k}^{*})|)=\pm\bm{\delta}$, we expand  $\bm{n}(\bm{k},t^{*})$ to the lowest order in the infinitesimal variable $q$ at the points $(k_{x},k_{y})=(k_{x}^{*}+q\cos\alpha,k_{y}^{*}+q\sin\alpha)$ as follows
\begin{equation}\label{d1}
	\begin{aligned}
		\bm{n}(\bm{k},t^{*})=&\bm{\delta}(\pm 1+\partial_{k_{x}}n_{\delta}\big|_{\bm{k}^{*}}q\cos\alpha+\partial_{k_{y}} n_{\delta}\big|_{\bm{k}^{*}}q\sin\alpha)\\&+\bm{x'}(A~q\cos\alpha+B~q\sin\alpha)+\bm{y'}(G~q\cos\alpha+H~q\sin\alpha),
	\end{aligned}
\end{equation}
here $A=\partial_{k_{x}}n_{x'}\big|_{\bm{k}^{*}}$, $B=\partial_{k_{y}}n_{x'}\big|_{\bm{k}^{*}}$, $G=\partial_{k_{x}}n_{y'}\big|_{\bm{k}^{*}}$, $H=\partial_{k_{y}}n_{y'}\big|_{\bm{k}^{*}}$. Making the substitution of the independent variable as $\zeta=e^{i\alpha}$, the vorticity can be defined as the phase accumulation around $\bm{k}^{*}$ at time $t^{*}$:
\begin{equation}\label{d2}
	\begin{aligned}
		&\nu(\bm{k}^{*},t^{*})=\lim_{q\to 0}\frac{1}{2\pi}\int_{0}^{2\pi}\frac{\partial\phi^{\prime}}{\partial\alpha}d\alpha=
		\frac{1}{2\pi}\oint_{|\zeta|=1}\mathcal{G}(\zeta)d\zeta,
	\end{aligned}
\end{equation}
where the azimuthal phase is described as $\phi^{\prime}(\bm{k},t^{*})\equiv\arctan[n_{y'}(\bm{k},t^{*})/n_{x'}(\bm{k},t^{*})] =Im[ln(A\cos\alpha+B\sin\alpha+iG\cos\alpha+iH\sin\alpha)]$, and thus 
\begin{equation}\label{d3}
	\mathcal{G}(\zeta)=\frac{4i(AH-BG)\zeta}{B^{2}(\zeta^{2}-1)^{2}-A^{2}(\zeta^{2}+1)^{2}+2iAB(\zeta^{4}-1)+[H(\zeta^{2}-1)+iG(\zeta^{2}+1)]^{2}}.
\end{equation}
By the residue theorem, the value of a contour integral in the complex plane depends only on the summation of the residues at the isolated singularities inside the contour. 
The isolated singularities of $\mathcal{G}(\zeta)$ are $\zeta_{1}=-\zeta_{2}=-\sqrt{\frac{-A-iB+iG-H}{A-iB-iG-H}}$ and $\zeta_{3}=-\zeta_{4}=-\sqrt{\frac{-A-iB-iG+H}{A-iB+iG+H}}$, so Eq.(\ref{d2}) can be further written as
\begin{equation}\label{d4}
	\nu=i\sum\limits_{|\zeta_{i}|<1}\mathrm{Res}[\mathcal{G}(\zeta_{i})],
\end{equation}
with the residues $\mathrm{Res}[\mathcal{G}(\zeta_{1})]=\mathrm{Res}[\mathcal{G}(\zeta_{2})]=\frac{i}{2}$ and
$\mathrm{Res}[\mathcal{G}(\zeta_{3})]=\mathrm{Res}[\mathcal{G}(\zeta_{4})]=-\frac{i}{2}$.  Three cases exist:

If $AH>BG$, only $\zeta_{3}$ and $\zeta_{4}$ are located in the unit circle $|\zeta|=1$;

If $AH<BG$, only $\zeta_{1}$ and $\zeta_{2}$ are located in unit the circle $|\zeta|=1$;	

If $AH=BG$, all the singularities are on the boundary of the circle. 	

Summarizing the above analysis, we can present the vorticity in the form of a sign function of a Jacobian determinant:
\begin{equation}\label{d5}
	\nu(\bm{k}^{*},t^{*})=\mathrm{sgn}(AH-BG)=\mathrm{sgn}\Big[\mathrm{det}\Big[\begin{bmatrix}\partial_{k_{x}}n_{x'}(\bm{k},t^{*}) & \partial_{k_{x}}n_{y'}(\bm{k},t^{*})\\\partial_{k_{y}}n_{x'}(\bm{k},t^{*})&\partial_{k_{y}}n_{y'}(\bm{k},t^{*})\\\end{bmatrix}\Big]\Big|_{\bm{k}^{*}}\Big],
\end{equation}
which is mirror invariant in the $\bm{\delta}$ direction, i.e, $\nu(n_{\delta},\bm{k}^{*},t^{*})=\nu(-n_{\delta},\bm{k}^{*},t^{*})$.

\section{Appendix E:\,\,\,Information of vortices contained in Berry curvatures}
\begin{figure}
	\scalebox{0.6}[0.6]{\includegraphics{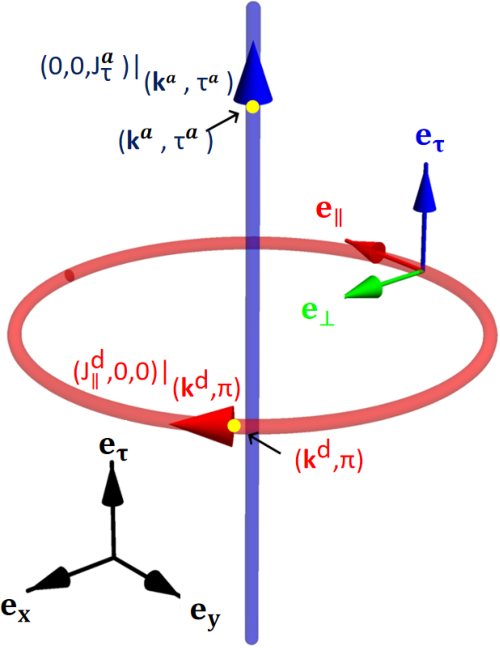}}
	\caption{\label{figs2}(Color online) Schematic configuration  of an orthogonal link consisting of the preimage loop in the $\tau=\pi$ plane and the preimage straight line along the time direction. The linking number $Lk=-1$ under the right-hand system. The local reference frame on the preimage loop has been well oriented, here $\mathrm{sgn}[J_{\varparallel}^{d}(\bm{k}^{d},\pi)]=-1$.}
\end{figure}

Under the parallel quench, the $\bm{k}$-independent quenching direction $\pm\bm{\delta}$ plays a crucial role in associating the Berry curvature with the vortex. 
First, Eq.~(\ref{a1}) produces the Berry curvature on the preimage straight line at a fixed point $\bm{k}^{s}(\in\bm{k}^{\ast})$, that is,  $\bm{J}^{s}(\bm{k}^{s},\tau^{s})=[0,0,J_{\tau}^{s}(\bm{k}^{s},\tau^{s})]$, and concretely
\begin{equation}\label{e1}
	J_{\tau}^{s}(\bm{k}^{s},\tau^{s})=\frac{1}{2}\bm{n}(\bm{k}^{s})\cdot[\partial_{k_{x}}\bm{n}(\bm{k},t^{s})\times\partial_{k_{y}}\bm{n}(\bm{k},t^{s})]\big|_{\bm{k}^{s}}=\frac{1}{2}n_{\delta}(\bm{k}^{s})\mathrm{det}\Big[\begin{bmatrix}\partial_{k_{x}}n_{x'}(\bm{k},t^{s}) & \partial_{k_{x}}n_{y'}(\bm{k},t^{s})\\\partial_{k_{y}}n_{x'}(\bm{k},t^{s})&\partial_{k_{y}}n_{y'}(\bm{k},t^{s})\\\end{bmatrix}\Big]\Big|_{\bm{k}^{s}}. 
\end{equation}

Combining with Eq.~(\ref{d5}), we obtain
\begin{equation}\label{e2}
	\mathrm{sgn}[J_{\tau}^{s}(\bm{k}^{s},\tau^{s})]=n_{\delta}(\bm{k}^{s})\nu^{s}(\bm{k}^{s},t^{s})
\end{equation}
with $n_{\delta}(\bm{k}^{s})$ being limited to $\pm1$, which indicates that the nonvanishing static vortex will ensure the nonzero Berry curvature. Inserting Eq.~(\ref{c3}) into Eq.~(\ref{d5}) yields the vorticity of the static vortex at any time $t^{s}$:
\begin{equation}\label{e3}
	\nu^{s}(\bm{k}^{s},t^{s})=\mathrm{sgn}[\partial_{k_{x}}n_{x'}(\bm{k},t^{s})\partial_{k_{y}}n_{y'}(\bm{k},t^{s})-\partial_{k_{x}}n_{y'}(\bm{k},t^{s})\partial_{k_{y}}n_{x'}(\bm{k},t^{s})]|_{\bm{k}^{s}}=\mathrm{sgn}[S(\bm{k}^{s})]\nu^{s}(\bm{k}^{s},0),
\end{equation}
here we always have 
\begin{equation}\label{e4} 
	\,\,S(\bm{k}^{s})\equiv[H^{f}_{\delta}(\bm{k}^{s})]^{2}+2[\cos(|\bm{H}^{f}(\bm{k}^{s})|t^{s})-1][H^{f}_{\delta}(\bm{k}^{s})-H^{i}_{\delta}(\bm{k}^{s})]H^{i}_{\delta} (\bm{k}^{s})\geq0,
\end{equation}\label{e5}
in view of the basic inequalities as follows
\begin{equation}
	\begin{aligned}
		&\cos(|\bm{H}^{f}(\bm{k}^{s})|t^{s})\left\{
		\begin{aligned}
			\geq-\frac{1}{2}\left(\frac{H^{f}_{\delta}(\bm{k}^{s})-H^{i}_{\delta}(\bm{k}^{s})}{H^{i}_{\delta}(\bm{k}^{s})}+\frac{H^{i}_{\delta}(\bm{k}^{s})}{H^{f}_{\delta}(\bm{k}^{s})-H^{i}_{\delta}(\bm{k}^{s})}\right):\,\,\,\,\,\,for\,\, [H^{f}_{\delta}(\bm{k}^{s})-H^{i}_{\delta}(\bm{k}^{s})]H^{i}_{\delta}(\bm{k}^{s})>0, \\
			\leq-\frac{1}{2}\left(\frac{H^{f}_{\delta}(\bm{k}^{s})-H^{i}_{\delta}(\bm{k}^{s})}{H^{i}_{\delta}(\bm{k}^{s})}+\frac{H^{i}_{\delta}(\bm{k}^{s})}{H^{f}_{\delta}(\bm{k}^{s})-H^{i}_{\delta}(\bm{k}^{s})}\right):\,\,\,\,\,\,for\,\, [H^{f}_{\delta}(\bm{k}^{s})-H^{i}_{\delta}(\bm{k}^{s})]H^{i}_{\delta}(\bm{k}^{s})<0.
		\end{aligned}
		\right.\\
	\end{aligned}
\end{equation}
The nonnegative function $S(\bm{k}^{s})$ implies that the static vorticity can be regarded as a conserved quantity, despite the zero vorticity is possible at some time instants when taking ``$=$'' in above inequalities.

Next, we will gain insights into the Berry curvature of the preimage loop  $\bm{J}^{d}(\bm{k}^{d},\pi)$ ($\bm{k}^{d}\in\bm{k}^{\ast}$) lying on the $\tau=\pi$ plane.
For this purpose, it is proper to convert the global reference frame  $(\bm{e}_{x},\bm{e}_{y},\bm{e}_{\tau})$ in the 3D space to the local frame $(\bm{e}_{\varparallel},\bm{e}_{\bot},\bm{e}_{\tau})$ on the preimage loop as depicted in Fig.\ref{figs2},
where  the counter-clockwise direction is set to be  positive in the top view.
Then we have $\bm{J}^{d}(\bm{k}^{d},\pi)=[J_{\varparallel}^{d}(\bm{k}^{d},\pi),0,0]$ with
\begin{equation}\label{e6} J_{\varparallel}^{d}(\bm{k}^{d},\pi)=\frac{1}{2}[\bm{n}\cdot(\partial_{k_{\bot}}\bm{n}\times\partial_{\tau}\bm{n})]\big|_{(\bm{k}^{d},\pi)}=\frac{1}{2}n_{\delta}(\bm{k}^{d},\pi)[\partial_{k_{\bot}}n_{\varparallel}(\bm{k},\tau)\partial_{\tau}n_{\bot}(\bm{k},\tau)-\partial_{k_{\bot}}n_{\bot}(\bm{k},\tau)\partial_{\tau}n_{\varparallel}(\bm{k},\tau)]\big|_{(\bm{k}^{d},\pi)},
\end{equation}
here $n_{\delta}(\bm{k}^{d},\pi)$ is always equal to $-\mathrm{sgn}[f(\bm{k}^{d})]$.  $\pi=|\bm{H}^{f}(\bm{k}^{d})|t^{d}$ schedules the first time slot for the evolution of the dynamical vortex.
Utilizing Eq.~(\ref{d5}), noting that $\tau=|\bm{H}^{f}(\bm{k})|t$ also includes $k_{\bot}$ and $k_{\varparallel}$, the vorticity of the dynamical vortex can be straightforwardly derived as
\begin{equation}\label{e7}
	\begin{aligned}
		\nu^{d}(\bm{k}^{d},t^{d})=&\mathrm{sgn}\big\{[\partial_{k_{\varparallel}}n_{\varparallel}(\bm{k},t^{d})\partial_{k_{\perp}}n_{\perp}(\bm{k},t^{d})-\partial_{k_{\perp}}n_{\varparallel}(\bm{k},t^{d})\partial_{k_{\varparallel}}n_{\perp}(\bm{k},t^{d})]\big|_{\bm{k}^{d}}\big\}\\=&\mathrm{sgn}\big\{[(\partial_{k_{\varparallel}}n_{\varparallel}(\bm{k},\tau)+\partial_{\tau}n_{\varparallel}(\bm{k},\tau)\partial_{k_{\varparallel}}|\bm{H}^{f}(\bm{k})| t^{d})(\partial_{k_{\perp}}n_{\perp}(\bm{k},\tau)+\partial_{\tau}n_{\perp}(\bm{k},\tau)\cdot\partial_{k_{\perp}}|\bm{H}^{f}(\bm{k})| t^{d})\\&-(\partial_{k_{\perp}}n_{\varparallel}(\bm{k},\tau)+\partial_{\tau}n_{\varparallel}(\bm{k},\tau)\partial_{k_{\perp}}|\bm{H}^{f}(\bm{k})| t^{d})(\partial_{k_{\varparallel}}n_{\perp}(\bm{k},\tau)+\partial_{\tau}n_{\perp}(\bm{k},\tau)\partial_{k_{\varparallel}}|\bm{H}^{f}(\bm{k})| t^{d})]\big|_{(\bm{k}^{d},\pi)}\big\}\\=&\mathrm{sgn}\big\{[\partial_{k_{\varparallel}}|\bm{H}^{f}(\bm{k})| t^{d}\big(\partial_{k_{\perp}}n_{\perp}(\bm{k},\tau)\partial_{\tau}n_{\varparallel}(\bm{k},\tau)-\partial_{k_{\perp}}n_{\varparallel}(\bm{k},\tau)\partial_{\tau}n_{\perp}(\bm{k},\tau)\big)]\big|_{(\bm{k}^{d},\pi)}\big\}\\=&\mathrm{sgn}[f(\bm{k}^{d})]\mathrm{sgn}[J_{\varparallel}^{d}(\bm{k}^{d},\pi)]\mathrm{sgn}[\partial_{k_{\varparallel}}|\bm{H}^{f}|\big|_{\bm{k}^{d}}],
	\end{aligned}
\end{equation}
where Eq.~(\ref{e6}) and the fact that $\partial_{k_{\varparallel}}n_{\varparallel}(\bm{k},\pi)=\partial_{k_{\varparallel}}n_{\perp}(\bm{k},\pi)=0$ have been considered.  Finally, the winding direction of the preimage loops is given by
\begin{equation}\label{e7}
	\begin{aligned}
		\omega^{J}\equiv\mathrm{sgn}[J_{\varparallel}^{d}(\bm{k}^{d},\pi)]=-\mathrm{sgn}[f(\bm{k}^{d})]\nu^{d}(\bm{k}^{d},\pi)\omega^{d}(\bm{k}^{d},\pi),
	\end{aligned}
\end{equation}
here $\omega^{d}(\bm{k}^{d},\pi)=-\mathrm{sgn}(\partial_{k_{\varparallel}}|\bm{H}^{f}|\big|_{\bm{k}^{d}})$ called `chirality' specifies the moving direction of the vortex on the trajectory.

\section{Appendix F:\,\,\,Ascribing the change of the Chern number to the magnetic field flipping at fixed points}
Our conclusions in this section stem from the primary fact that, for any given Hamiltonian $\mathcal{H}(\bm{k})$, its Chern numbers can be expressed as the sum of vorticities. Without loss of the generality, we suppose that the ground state of the Hamiltonian $\mathcal{H}(\bm{k})$
is parameterized as $|\psi(\bm{k})\rangle=(
-\sin[\theta(\bm{k})/2]e^{-i\phi(\bm{k})},
\cos[\theta(\bm{k})/2] 
)^{\mathrm{T}}$ and $L$ zero points with $\bm{h}(\bm{k}^{s})=[\sin\theta(\bm{k}^{s})\cos\phi(\bm{k}^{s}),\sin\theta(\bm{k}^{s})\sin\phi(\bm{k}^{s}),\cos\theta(\bm{k}^{s})]=(0,0,\pm1)$ exist, and then expand $\bm{k}_{l}$ near the $l$th zero point to the lowest order in $q_{l}$ as  $\bm{k}_{l}=\bm{k}_{l}^{s}+ q_{l}(\cos\alpha_{l},\sin\alpha_{l})$. By the Stokes Theorem, the Chern number of the lower band can be written as the sum of the line integrals of the Berry connection around each zero point $\bm{k}^{s}_{l}$:
\begin{equation}\label{f2}
	\begin{aligned}
		C&=\frac{1}{2\pi}\iint_{BZ}dk_{x}dk_{y}(\partial_{k_{x}}A_{k_{y}}-\partial_{k_{y}}A_{k_{x}})\\&=-\frac{1}{2\pi}\ointctrclockwise\limits_{\partial BZ}(A_{k_{x}}dk_{x}+A_{k_{y}}dk_{y})=-\frac{1}{2\pi}\sum\limits_{l=1}^{	L}\Big[\lim\limits_{q_{l}\rightarrow0}\ointctrclockwise\limits_{l}(A_{\alpha_{l}}d\alpha_{l}+A_{q_{l}}dq_{l}) \Big]\\&=\frac{1}{2\pi}\sum\limits_{l=1}^{	L}\Big[\lim\limits_{q_{l}\rightarrow0}\ointctrclockwise\limits_{l}\frac{1-\cos\theta(\bm{k}_{l})}{2}d\phi(\bm{k}_{l}) \Big]
		=\sum\limits_{l=1}^{	L}\frac{1-\cos\theta(\bm{k}_{l}^{s})}{2}\nu^{s}(\bm{k}_{l}^{s}),
	\end{aligned}
\end{equation}
here the boundary of the Brillouin Zone $\partial BZ$ consists of $L$ circles with infinitesimal radius centered at $\bm{k}^{s}_{l}$, where the vorticity  $\nu(\bm{k}^{s}_{l})$ will be counted in the Chern number once $\theta(\bm{k}^{s}_{l})=\pi$.  $A_{\mu}=-i\langle\psi(\bm{k})|\partial_{\mu}|\psi(\bm{k})\rangle$ is the Berry connection. In particular, $A_{\alpha_{l}}= \frac{\cos\theta(\bm{k}_{l})-1}{2}\partial_{\alpha_{l}}\phi(\bm{k}_{l})$. Note that, under the gauge transformation $|\psi(\bm{k})\rangle\rightarrow e^{i\varphi(\bm{k})}|\psi(\bm{k})\rangle$ forming a $U(1)$ fiber bundle, although  $A_{\alpha_{l}}$ is transformed to be $A_{\alpha_{l}}+\partial_{\alpha_{l}}\varphi$, the vorticity $\nu(\bm{k}^{s}_{l})$ of phase $\phi(\equiv\arctan[h_{y}/h_{x}])$
is manifestly gauge independent due to $\lim\limits_{q_{l}\rightarrow0}\ointctrclockwise\limits_{l}d\varphi=0$.

Thanks to the parallel quench, for all $\bm{k}$,
in the new set of axes $(\bm{x'},\bm{y'},\bm{\delta})$ only
the term $h_{\delta}$ has changed under the quench process from $\bm{h}^{i}$ to $\bm{h}^{f}$, while the terms $h_{x'}$ and $h_{y'}$ remain constant, thus keeping the azimuthal phase $\phi^{\prime}$ fixed. Moreover, the zero points with $h_{\delta}(\bm{k}_{l}^{s})=\pm1$ also become the fixed points, which mean that at these points the pre-quench and post-quench magnetic fields can only align parallel or antiparallel to each other.   For the pre- (or post-) quench magnetic field $\bm{H}^{i, f}=(H_{x'}^{i,f},H_{y'}^{i,f},H_{\delta}^{i,f})=|\bm{H}^{i,f}|(\sin\theta^{\prime i,f}\cos\phi^{\prime},\sin\theta^{\prime i,f}\sin\phi^{\prime},\cos\theta^{\prime i,f})$, suppose that $M$ fixed points are antiparallel (called ``singular points''), i.e.,
\begin{equation}
	\bm{k}^{a}\in\{\bm{k}^{s}|\bm{h}^{i}(\bm{k}^{s})\cdot\bm{h}^{f}(\bm{k}^{s})=-1\},
\end{equation}
considering Eq.~(\ref{c1}) and Eq.~(\ref{c2}) gains  $[H_{\delta}^{i}(\bm{k}^{a})+f(\bm{k}^{a})]/|H_{\delta}^{i}(\bm{k}^{a})+f(\bm{k}^{a})|=-H_{\delta}^{i}(\bm{k}^{a})/|H_{\delta}^{i}(\bm{k}^{a})|$, so $H_{\delta}^{i}(\bm{k}^{a})$ and $f(\bm{k}^{a})$ must own the different signs, i.e., $\mathrm{sgn}[f(\bm{k}^{a})]=-h^{i}_{\delta}(\bm{k}^{a})= n_{\delta}(\bm{k}^{a})$.
Then, we obtain the difference of the Chern number according to Eq.~(\ref{f2}):
\begin{equation}\label{f3}
	C_{f}-C_{i}=\sum\limits_{l=1}^{L}[\frac{\cos\theta^{\prime i}(\bm{k}_{l}^{s})-\cos\theta^{\prime f}(\bm{k}_{l}^{s})}{2}\nu^{s}(\bm{k}_{l}^{s})]=-\sum\limits_{l=1}^{M}\mathrm{sgn}[f(\bm{k}_{l}^{a})]\nu^{s}(\bm{k}_{l}^{a}),
\end{equation}
definitely indicating that the magnetic field flipping always changes the Chern number by taking the relevant vortex into account. In Appendix C we have pointed out the horizontal preimages can only belong to $n_{\delta}(\bm{k}^{d},\pi)=-\mathrm{sgn}[f(\bm{k}^{d})]$. To this extent, we further require the preimage loop to surround the vertical preimages at the magnetic field flipping points $\bm{k}^{a}$ by choosing a proper Brillouin zone. This is always achievable.
The orthogonal links can be constructed more explicitly  as follows

\begin{equation}\label{f4}
	\begin{aligned}
		& orthogonal\,\, links=\left\{
		\begin{aligned}
			vertical\,\, lines\,\,(\bm{k}^{a}\in\bm{k}^{s}\in\bm{k}^{*},\tau^{a}\in[0, 2\pi)):\,\,\,\,\,\, &n_{\delta}(\bm{k}^{a},\tau^{a})=\mathrm{sgn}[f(\bm{k}^{a})],\\
			horizontal \,\,loops\,\,\,\,\,\,\,\,\,\,\,\,\,\,\,\,\,\,\,\,\,\,\,\,\,\,(\bm{k}^{d}\in\bm{k}^{*},\tau^{d}=\pi): \,\,\,\,\,\,&n_{\delta}(\bm{k}^{d},\tau^{d})=-\mathrm{sgn}[f(\bm{k}^{d})].
		\end{aligned}
		\right.\\
	\end{aligned}
\end{equation}

\begin{figure}
	\scalebox{1.2}[1.18]{\includegraphics{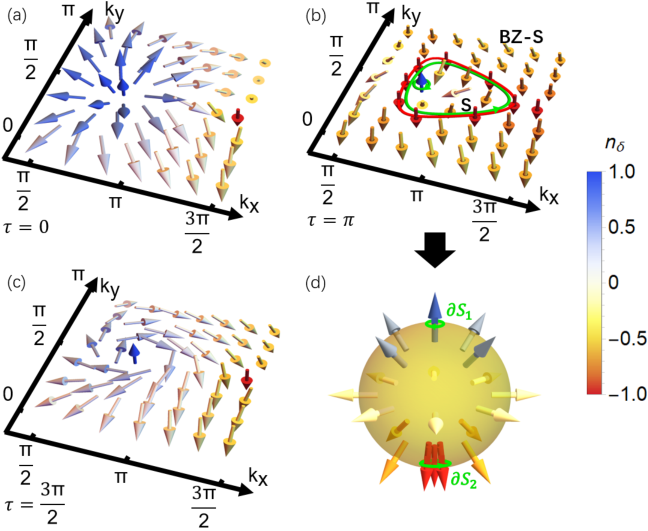}}
	\caption{\label{figs3}(Color online) Topological spin textures $\bm{n}(\bm{k},\tau)$ for the quench from $C_{i}=1$ to $C_{f}=0$ phase at the different rescaled time: $\tau=0$ (a), $\tau=\pi$ (b), $\tau=3\pi/2$ (c). The quenching parameters $(m^{i}_{x},m^{i}_{z},m^{f}_{x},m^{f}_{z})=(0.5,1,0.8,1.8)$ coincide with the Fig.\ref{fig2}(b) in the main text.  A quarter of the first Brillouin zone $\{k_{x}\in[\frac{\pi}{2},\frac{3\pi}{2}),k_{y}\in[0,\pi)\}$ is enough to present the forming process of the skyrmion with $SK=1$ meeting $SK=C_{i}-C_{f}$.
		The arrows are colored by the $n_{\delta}$ components of corresponding vectors $\bm{n}$. (d) The skyrmionic spin pattern in the zone $S$ morphed into a Bloch sphere. The preimages of $n_{\delta}(\bm{k}^{d},\pi)=-1$ are located on the red loop, while the green oriented loops mark the path and the direction of line integrals.}
\end{figure}

A new enlightenment given by Eq.~(\ref{f4}) is that the Bloch vector $\bm{n}(\bm{k},\tau)$ in the $\tau=\pi$ plane will be able to form various skyrmions with spin swirling textures \cite{skyrme1962unified}. 
To be concrete, we characterize the spin textures in Fig.\ref{figs3} (a), (b) and (c) for different $\tau$. The relevant values are taken from Fig.\ref{fig2}(b) in the main text. It is obvious from Fig.\ref{figs3} (b) that the preimage loop encloses a skyrmion-like configuration in the zone $S$, its Pontryagin number \cite{braun2012topological,pontrjagin1941classification} can be computed as 
\begin{equation}\label{f6}
	\begin{aligned}
		\\
		SK&=\frac{1}{4\pi}\iint_{S}dk_{x}dk_{y}\bm{n}(\bm{k},\pi)\cdot[\partial_{k_{x}}\bm{n}(\bm{k},\pi)\times\partial_{k_{y}}\bm{n}(\bm{k},\pi)]\\&=\frac{1}{2\pi}\iint_{S}dk_{x}dk_{y}[\partial_{k_{x}}A_{k_{y}}(\bm{k},\pi)-\partial_{k_{y}}A_{k_{x}}(\bm{k},\pi)]\\&=\frac{1}{2\pi}\ointctrclockwise\limits_{\partial S_{2}}[A_{k_{x}}(\bm{k},\pi)dk_{x}+A_{k_{y}}(\bm{k},\pi)dk_{y}]-\frac{1}{2\pi}\ointctrclockwise\limits_{\partial S_{1}}[A_{k_{x}}(\bm{k},\pi)dk_{x}+A_{k_{y}}(\bm{k},\pi)dk_{y}]\\&=[\frac{\cos\theta^{\prime}(\bm{k}^{d},\pi)-1}{2}\overline{\nu}(\bm{k}^{d})]-[\frac{\cos\theta^{\prime}(\bm{k}^{a},\pi)-1}{2}\nu^{s}(\bm{k}^{a})]=\frac{1-\cos\theta^{\prime}(\bm{k}^{a},0)}{2}\nu^{s}(\bm{k}^{a})=\nu^{s}(\bm{k}^{a}),
	\end{aligned}
\end{equation}
where the area integral over the zone $S$ has been converted into the line integrals along the edges $\partial S_{1}$ and $\partial S_{2}$ [see the green loops in Fig.\ref{figs3} (b) and (d)]. Note that the nonvanishing vorticity at the fixed point (the skyrmion core) does not change with time
as indicated in Eq.~(\ref{e4}). $n_{\delta}(\bm{k}^{a},\tau)=1$ means $\theta^{\prime}(\bm{k}^{a},\pi)=\theta^{\prime}(\bm{k}^{a},0)=\pi$ due to $\bm{h}^{i}(\bm{k}^{a},\tau)=-\bm{n}(\bm{k}^{a},\tau)$, and $\nu^{s}(\bm{k}^{a})=1$ actually makes a skyrmion. The reason why the first term in the fourth line disappeared is that the edges of the skyrmion can be glued  together through a continuous topological deformation and, consequently, all the points with  $n_{\delta}(\bm{k}^{d},\pi)=-1$ merge into the south pole, of which the vorticity $\overline{\nu}(\bm{k}^{d})$ will not be counted as $\theta^{\prime}(\bm{k}^{d},\pi)=0$. In addition, because the unitary evolution governed by $U(\bm{k},\tau)\equiv e^{-i \frac{\tau}{2} \bm{h}^{f}(\bm{k})\cdot\bm{\sigma}}$ preserves the Chern number (but changes the Pontryagin density) \cite{d2015dynamical}, we further have $C_{i}=C_{\tau=\pi}=SK+C_{BZ-S}$ with $C_{BZ-S}=0$ being offered by the rest part of the $\tau=\pi$ plane $BZ-S$ [see Fig.\ref{figs3} (b)].
The biskyrmion as mentioned in the main text can also be treated in a similar way. In a word, whether they are skyrmions,  anti-skyrmions or biskyrmions, their total Pontryagin numbers in the $\tau=\pi$ plane will equal to the difference of the Chern number between $\mathcal{H}^{i}(\bm{k})$ and $\mathcal{H}^{f}(\bm{k})$ in terms of Eq.~(\ref{f3}).

\section{Appendix G:\,\,\,Proof for $\omega^{J}=-1$}
In this section, we will prove that the preimage loop wraps around the antiparallel fixed point $\bm{k}^{a}$ always clockwise, i.e.,
\begin{equation}\label{g1}
	\omega^{J}\equiv\mathrm{sgn}[J_{\varparallel}^{d}(\bm{k}^{d},\pi)]=-1.
\end{equation} 

First, from Eq.~(\ref{a1}) the Berry curvature $J_{\varparallel}^{d}(\bm{k}^{d},\pi)$ on the preimage loop can be calculated as 
\begin{equation}\label{g2}
	J_{\varparallel}^{d}(\bm{k}^{d},\pi)=\frac{1}{2}\bm{n}\cdot(\partial_{k_{\perp}}\bm{n}\times\partial_{\tau}\bm{n})\big|_{(\bm{k}^{d},\pi)}=\frac{1}{2}\bm{n}\cdot[(\partial_{k_{\perp}}\bm{n}\cdot\bm{n})\bm{h}^{f}-(\partial_{k_{\perp}}\bm{n}\cdot\bm{h}^{f})\bm{n}]\big|_{(\bm{k}^{d},\pi)}=-\frac{1}{2}\partial_{k_{\perp}}\bm{n}\cdot\bm{h}^{f}\big|_{(\bm{k}^{d},\pi)},
\end{equation}
here we have made use of the fact that $\partial_{\tau}\bm{n}=\bm{h}^{f}\times\bm{n}$ and $\partial_{k_{\perp}}\bm{n}\cdot\bm{n}=0$. Since $n_{x^{\prime}}(\bm{k}^{d},\pi)=n_{y^{\prime}}(\bm{k}^{d},\pi)=0$ and $n_{\delta}(\bm{k}^{d},\pi)=-\mathrm{sgn}[f(\bm{k}^{d})]$, we have $\bm{n}(\bm{k}^{d},\pi)\cdot\partial_{k_{\perp}}\bm{n}|_{(\bm{k}^{d},\pi)}=n_{\delta}(\bm{k}^{d},\pi)\partial_{k_{\perp}}n_{\delta}(\bm{k})|_{(\bm{k}^{d},\pi)}=-\mathrm{sgn}[f(\bm{k}^{d})]\partial_{k_{\perp}}n_{\delta}(\bm{k})|_{(\bm{k}^{d},\pi)}=0$. Further, we can get
\begin{equation}\label{g3}
	\begin{aligned}
		\partial_{k_{\perp}}\bm{n}(\bm{k})\cdot\bm{h}^{f}(\bm{k})|_{(\bm{k}^{d},\pi)}=\partial_{k_{\perp}}\bar{\bm{n}}(\bm{k})\cdot\bar{\bm{h}}^{f}(\bm{k})|_{(\bm{k}^{d},\pi)}=\partial_{k_{\perp}}[\bar{\bm{n}}(\bm{k})\cdot\bar{\bm{h}}^{f}(\bm{k})]|_{(\bm{k}^{d},\pi)}=\partial_{k_{\perp}}g(\bm{k})|_{\bm{k}^{d}},
	\end{aligned}
\end{equation} 
here the scalar- product function $g(\bm{k})\equiv\bar{\bm{n}}(\bm{k},\pi)\cdot\bar{\bm{h}}^{f}(\bm{k},\pi)\equiv n_{x^{\prime}}(\bm{k},\pi)h^{f}_{x^{\prime}}(\bm{k},\pi)+n_{y^{\prime}}(\bm{k},\pi)h^{f}_{y^{\prime}}(\bm{k},\pi)$ describes a smooth surface. Next, the proof of Eq.~(\ref{g1}) is reduced to
the proof of $\partial_{k_{\perp}}g(\bm{k})|_{\bm{k}^{d}}>0$.

\begin{figure}
	\scalebox{1}[0.8]{\includegraphics{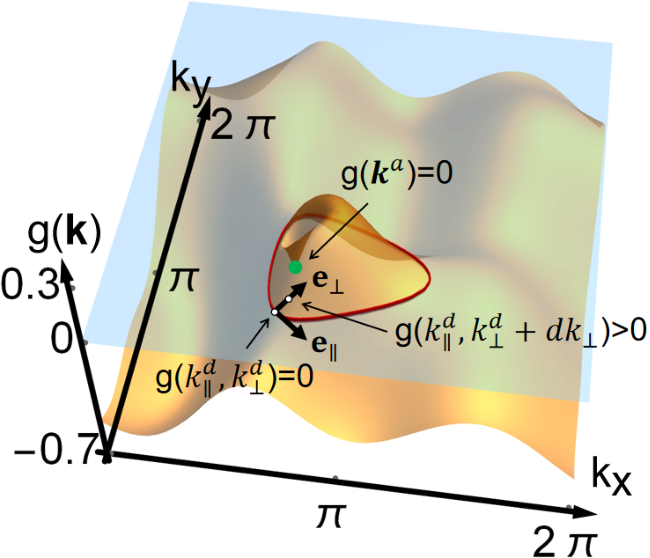}}
	\caption{\label{figs4}(Color online) The scalar-product function $g(\bm{k})$ with same quench parameters as that in Fig.\ref{figs3}.
		The winding direction of the preimage (red) loop is decided by the properties of this function at $\bm{k}^{a}$. The second partial derivative test for $g(\bm{k}^{a})$ (green dot) results in  $\partial_{k_{\perp}}g(\bm{k})|_{\bm{k}^{d}}=\lim_{dk_{\perp}\to 0^{+}}\frac{g(k_{\varparallel}^{d},k_{\perp}^{d}+dk_{\perp})-g(k_{\varparallel}^{d},k_{\perp}^{d})}{dk_{\perp}}>0$.}
\end{figure}

Note that $g(\bm{k})=0$ establishes only at $\bm{k}^{s}$ and $\bm{k}^{d}$. The preimage loops satisfying $g(\bm{k}^{d})=0$ divide the Brillouin zone into different patches. Actually, $g(\bm{k}^{a})(\bm{k}^{a}\in\bm{k}^{s})$ is a local minimum, thus we can always make a smooth path with $g(\bm{k})>0$ connect the point $\bm{k}^{a}$ to an arbitrary point $\bm{k}^{d}$ on the preimage loop encircling this $\bm{k}^{a}$. Due to the fact that the preimage loop must own a direction, $\partial_{k_{\perp}}g(\bm{k})|_{\bm{k}^{d}}>0$ can be proved (note the direction of $k_{\perp}$ in Fig.\ref{figs4}).

$\ddagger$ Proof that $g(\bm{k}^{a})$ is a local minimum:

Let $g(x,y)$ be a differentiable function of two variables and own the continuous second partial derivatives. Assume that the point $(a,b)$ satisfies $\partial_{x}g(x,y)|_{(a,b)}=\partial_{y}g(x,y)|_{(a,b)}=0$. Based on
the Hessian matrix 
\begin{equation}\label{g4}
	\begin{aligned} 
		D(x,y)=\begin{pmatrix}\frac{\partial^{2}g(x,y)}{\partial x^{2}} & \frac{\partial^{2}g(x,y)}{\partial x\partial y}\\\frac{\partial^{2}g(x,y)}{\partial y\partial x}&\frac{\partial^{2}g(x,y)}{\partial y^{2}}\\\end{pmatrix},
	\end{aligned}
\end{equation}

the method of the second partial derivative test can give the following criteria \cite{spring1985second,nerenberg1991second}:

1.If $\mathrm{det}[D(a,b)]>0$ and $\frac{\partial^{2}g(x,y)}{\partial x^{2}}\big|_{(a,b)}>0$, then $(a,b)$ is a local minimum of $g(x,y)$;

2.If $\mathrm{det}[D(a,b)]>0$ and $\frac{\partial^{2}g(x,y)}{\partial x^{2}}\big|_{(a,b)}<0$, then $(a,b)$ is a local maximum of $g(x,y)$;

3.If $\mathrm{det}[D(a,b)]<0$, then $(a,b)$ is a saddle point of $g(x,y)$;

4.If $\mathrm{det}[D(a,b)]=0$, then the second derivative test is inconclusive.

In fact, $\mathrm{det}[D(a,b)]$ is precisely the gaussian curvature of the smooth surface at $(a,b)$.  For the parallel quench process Eq.~(\ref{c1}), it is easy to check that  $g(\bm{k}^{s})=0$ and
$\frac{\partial g(\bm{k})}{\partial k_{x}}|_{\bm{k}^{s}}=\frac{\partial g(\bm{k})}{\partial k_{y}}|_{\bm{k}^{s}}=0$, as $H_{x'}(\bm{k}^{s})=H_{y'}(\bm{k}^{s})=0$ at all fixed points. In order to calculate its second partial derivative at $\bm{k}^{a}$, we write $g(\bm{k})$ in detail:

\begin{equation}\label{g5}
	\begin{aligned} 
		g(\bm{k})=&[h^{i}_{x'}-2(\bm{h}^{f}\cdot\bm{h}^{i})h^{f}_{x'}]h^{f}_{x'}+[h^{i}_{y'}-2(\bm{h}^{f}\cdot\bm{h}^{i})h^{f}_{y'}]h^{f}_{y'}\\=&\frac{H_{x'}^{2}(|H^{f}_{\delta}|^{2}-2H_{\delta}^{i}H_{\delta}^{f}-H_{x'}^{2}-H_{y'}^{2})}{|\bm{H}^{i}||\bm{H}^{f}|^{3}}+\frac{H_{y'}^{2}(|H^{f}_{\delta}|^{2}-2H_{\delta}^{i}H_{\delta}^{f}-H_{x'}^{2}-H_{y'}^{2})}{|\bm{H}^{i}||\bm{H}^{f}|^{3}}\\=&\frac{(H_{x'}^{2}+H_{y'}^{2})[(H_{\delta}^{f})^{2}-H_{x'}^{2}-H_{y'}^{2}-2H_{\delta}^{i}H_{\delta}^{f}]}{|\bm{H}^{i}||\bm{H}^{f}|^{3}},
	\end{aligned}
\end{equation}
which could determine $\bm{k}^{s}$ ($\bm{k}^{d}$) by setting the first (second) factor of the numerator of the last equality to zero. 
Given that the static vortex at magnetic field flipping point $\bm{k}^{a}$ exists, substituting Eq.~(\ref{g5}) into Eq.~(\ref{g4}) via a long but straightforward algebra yields
\begin{equation}\label{g6}
	\begin{aligned}
		\mathrm{det}[D(\bm{k}^{a})]&=\left [\frac{\partial^{2}g(\bm{k})}{\partial k_{x}^{2}}\frac{\partial^{2}g(\bm{k})}{\partial k_{y}^{2}}-\left(\frac{\partial^{2}g(\bm{k})}{\partial k_{x}\partial k_{y}}\right)^{2}\right]\bigg|_{\bm{k}^{a}}\\&=\left\{\frac{4(H_{\delta}^{f}-2H_{\delta}^{i})^{2}[(\partial_{k_{x}}H_{x})^{2}+(\partial_{k_{x}}H_{y})^{2}][(\partial_{k_{y}}H_{x})^{2}+(\partial_{k_{y}}H_{y})^{2}]}{(H_{\delta}^{i})^{2}(H_{\delta}^{f})^{4}}-\frac{4(H_{\delta}^{f}-2H_{\delta}^{i})^{2}(\partial_{k_{x}}H_{x}\partial_{k_{y}}H_{x}+\partial_{k_{x}}H_{y}\partial_{k_{y}}H_{y})^{2}}{(H_{\delta}^{i})^{2}(H_{\delta}^{f})^{4}}\right\}\bigg|_{\bm{k}^{a}}\\&=\left[\frac{4(H_{\delta}^{f}-2H_{\delta}^{i})^{2}(\partial_{k_{y}}H_{x'}\partial_{k_{x}}H_{y'}-\partial_{k_{x}}H_{x'}\partial_{k_{y}}H_{y'})^{2}}{(H_{\delta}^{i})^{2}(H_{\delta}^{f})^{4}}\right]\Bigg|_{\bm{k}^{a}}>0.
	\end{aligned}
\end{equation}

Noting that $H_{\delta}^{i}(\bm{k}^{a})$ and $H_{\delta}^{f}(\bm{k}^{a})$ are just opposite in sign, we also obtain
\begin{equation}\label{g7}
	\frac{\partial^{2}g(\bm{k})}{\partial k_{x}^{2}}\bigg|_{\bm{k}^{a}}=2\bigg(1-2\frac{H_{\delta}^{i}}{H_{\delta}^{f}}\bigg)\frac{(\partial_{k_{x}}H_{x'})^{2}+(\partial_{k_{x}}H_{y'})^{2}}{|H_{\delta}^{i}||H_{\delta}^{f}|}\bigg|_{\bm{k}^{a}}>0.
\end{equation}

This proof is achieved.$\ddagger$

\section{Appendix H:\,\,\,Proof for $Lk=C_{f}-C_{i}$}

In our setting, the horizontal preimages (if there are any) were required to encircle the $M$ magnetic field flipping points $\bm{k}_{l}^{a}$ ($l=1,2\ldots ,M$) in the Brillouin zone. 
Whether the horizontal preimages are closed or not inside the Brillouin zone, they can always be divided into $M$ subloops by adding the ordinary auxiliary lines in pairs. Each subloop plus the straight line at the antiparallel fixed point $\bm{k}_{l}^{a}$ constitutes a sublink with the linking number $Lk_{l}$, so that the total linking number $Lk=\sum\limits_{l=1}^{M}Lk_{l}$. For each sublink, we reiterate that $n_{\delta}(\bm{k}_{l}^{a})=\mathrm{sgn}[f(\bm{k}_{l}^{a})]$, and use the relations Eq.~(\ref{b2}), Eq.~(\ref{e2}) and Eq.~(\ref{g1}) to obtain
\begin{equation}\label{h1} 
	Lk_{l}=\mathrm{sgn}[J_{\tau}^{a}(\bm{k}_{l}^{a})]\omega^{J}_{l}=-n_{\delta}(\bm{k}_{l}^{a})\nu^{s}(\bm{k}_{l}^{a})=-\mathrm{sgn}[f(\bm{k}_{l}^{a})]\nu^{s}(\bm{k}_{l}^{a}).
\end{equation}
Armed with this, utilizing Eq.~(\ref{f3}) we eventually arrive at the conclusion
\begin{equation}
	Lk=\sum\limits_{l=1}^{M}Lk_{l}=-\sum\limits_{l=1}^{M}\mathrm{sgn}[f(\bm{k}_{l}^{a})]\nu^{s}(\bm{k}_{l}^{a})=C_{f}-C_{i}.
\end{equation}
\end{widetext}

\bibliographystyle{apsrev4-2}
\bibliography{Maintextreference}

\end{document}